\documentclass[useAMS,usenatbib]{mnras}

\usepackage[english,english]{babel}
\usepackage{amsmath}
\usepackage{amssymb,amsfonts,textcomp}
\usepackage{array}
\usepackage{comment}
\usepackage{supertabular}
\usepackage{multirow}
\usepackage{placeins} 
\usepackage{hhline}
\usepackage[usenames]{color}
\usepackage{soul}
\usepackage{textcomp}
\usepackage{graphicx}
\usepackage{float}
\usepackage{dblfloatfix}
\usepackage{dcolumn}
\usepackage{bm}
\usepackage{comment}
\usepackage{listings}
\usepackage{verbatim}
\usepackage{ulem}
 \usepackage{float}
\usepackage{times}
\usepackage[table,x11names]{xcolor}

\renewcommand{\deg}{{\rm deg}}

\newcommand{\dustgraincosmo}{{\small DUSTGRAIN-}\emph{cosmo}}
\newcommand{\dustgrainpathfinder}{{\small DUSTGRAIN-}\emph{pathfinder}}

\definecolor{Blue}{rgb}{0,0.2,0.65}
\definecolor{Green}{rgb}{0.2,0.55,0.35}
\definecolor{grey}{rgb}{0.75,0.75,0.75}
\definecolor{Orange}{rgb}{1.0,0.5,0.15}
\definecolor{brown}{rgb}{0.7,0.25,0.0}
\definecolor{Pink}{rgb}{1.0,0.5,0.5}
\definecolor{darkerred}{rgb}{0.8,0,0}
\definecolor{darkerblue}{rgb}{0,0,0.8}
\definecolor{darkergreen}{rgb}{0,0.5,0}
\definecolor{antiquefuchsia}{rgb}{0.57, 0.36, 0.51}
\definecolor{olive}{rgb}{0.51, 0.5, 0.345}
\definecolor{cyan}{rgb}{0.0, 0.72, 0.92}
\definecolor{tablecolour}{rgb}{1.0, 0.72, 0.77}

\begin{document}

\title[Convergence Cosmology]{Nuw CDM cosmology from the weak lensing convergence PDF}

\author[Boyle et al.]{
\parbox[t]{\textwidth}{Aoife Boyle$^{1}$\thanks{E-mail: boyle@iap.fr}, Cora Uhlemann$^{2}$, Oliver Friedrich$^{3,4}$, Alexandre Barthelemy$^{1}$,\\ 
Sandrine Codis$^{1,5}$, Francis Bernardeau$^{1,5}$, Carlo Giocoli$^{6,7}$, 
Marco Baldi$^{6,7,8}$}
\vspace*{10pt}\\
$^{1}$ CNRS \& Sorbonne Universit\'e, UMR 7095, Institut d'Astrophysique de Paris, 75014, Paris, France\\
$^{2}$ {School of Mathematics, Statistics and Physics, Newcastle University, Herschel Building, NE1 7RU Newcastle-upon-Tyne, United Kingdom}\\
$^3$ {Kavli Institute for Cosmology, University of Cambridge, CB3 0HA Cambridge, United Kingdom}\\
$^{4}$ {Churchill College, University of Cambridge, CB3 0DS Cambridge, United Kingdom}\\
$^{5}$ IPhT, DRF-INP, UMR 3680, CEA, Orme des Merisiers Bat 774, 91191 Gif-sur-Yvette, France \\
$^{6}$ INAF - Osservatorio di Astrofisica e
  Scienza dello Spazio di Bologna, via Piero Gobetti 93/3, I-40129 Bologna,
  Italy \\
$^{7}$ INFN - Sezione di Bologna, viale Berti Pichat 6/2, I-40127 Bologna, Italy\\
$^{8}$ Dipartimento di Fisica e Astronomia, Alma Mater Studiorum Universita di Bologna, via Piero Gobetti 93/2, I-40129 Bologna, Italy
}
\maketitle
\begin{abstract}
Pinning down the total neutrino mass and the dark energy equation of state is a key aim for upcoming galaxy surveys. Weak lensing is a unique probe of the total matter distribution whose non-Gaussian statistics can be quantified by the one-point probability distribution function (PDF) of the lensing convergence. We calculate the convergence PDF on mildly non-linear scales from first principles using large-deviation statistics, accounting for dark energy and the total neutrino mass. For the first time, we comprehensively validate the cosmology-dependence of the convergence PDF model against large suites of simulated lensing maps, demonstrating its percent-level precision and accuracy. We show that fast simulation codes can provide highly accurate covariance matrices, which can be combined with the theoretical PDF model to perform forecasts and eliminate the need for relying on expensive N-body simulations. Our theoretical model allows us to perform the first forecast for the convergence PDF that varies the full set of $\Lambda$CDM parameters. Our Fisher forecasts establish that the constraining power of the convergence PDF compares favourably to the two-point correlation function for a Euclid-like survey area at a single source redshift. When combined with a CMB prior from Planck, the PDF constrains both the neutrino mass $M_\nu$ and the dark energy equation of state $w_0$ more strongly than the two-point correlation function. 

 \end{abstract}

\normalem

\section{Introduction}
\label{sec:intro}

The path of light travelling from distant background galaxies is perturbed by matter density fluctuations along the line of sight. Distortions in the images of these background galaxies caused by weak gravitational lensing \citep[reviewed by][]{Kilbinger_2015} can be used to infer the convergence field and hence directly probe the projected matter density between the source galaxies and the observer. The subtle effects of weak gravitational lensing create a need for large statistics in order to constrain cosmological parameters, such as those acquired by the Dark Energy Survey \citep[DES,][]{Abbott_2018}, the Kilo-Degree Survey \citep[KiDS,][]{kids2020} and the Hyper Suprime Cam \citep[HSC,][]{HSC_2019}. The future promises even larger and more powerful weak lensing surveys such as Euclid \citep{Euclid, Euclid16} and the Rubin Observatory \citep{LSST}. 

The non-Gaussian evolution of the Universe at low redshifts and small scales implies that two-point statistics of the weak lensing field cannot capture all of the information available. Several previous works have demonstrated that the information content of convergence maps can be better captured when higher order statistics are included. Since it is difficult to find probes of higher order statistics whose cosmological dependence can be accurately derived, the bulk of analysis has relied on numerical simulations. 

\cite{Petri13} showed that adding Minkowski functionals or higher order moments to the weak lensing convergence power spectrum can provide meaningful improvements in the constraints on cosmological parameters. \cite{Petri_2016} showed that the non-Gaussian information contained in moments of the convergence field and convergence peak counts could also complement constraints from weak lensing tomography. \cite{Vicinanza2018} confirmed that moments of the convergence field can break some cosmological degeneracies originating in the shear power spectrum. \cite{Peel18} proposed the use of higher order weak lensing statistics to break degeneracies between massive neutrinos and modified gravity $f(R)$ models in the weak lensing power spectrum. \cite{davies2020constraining} recently examined the constraining power of weak lensing voids alone in a Rubin-Observatory-like survey, and showed that these statistics outperform the shear-shear correlation function as a probe. \cite{2020arXiv201007376M} investigated the constraining power of weak lensing aperture mass statistics, such as peaks, voids and the probability distribution function (PDF), also showing significant benefits over the shear two-point correlation function. \cite{Zuercher_2020} examined the benefits to constraints of peak counts, minimum counts or Minkowski functionals compared to the angular cosmic shear power spectrum, and confirmed that the non-Gaussian statistics could significantly enhance constraints in the ${\Omega_{\rm m}-\sigma_8}$ plane. A very recent study by \cite{harnoisderaps2020} found one of the tightest constraints on $S_8=\sigma_8\sqrt{\Omega_{\rm m}/0.3}$ in DES year-1 data based on a simulation-based combined peak count and correlation function analysis. \cite{munshi2020morphology} recently demonstrated how the Minkowski functionals of convergence maps can be inferred using a fitting function to the bispectrum.

The full convergence PDF includes more information than can be accessed through a small set of moments, with the additional benefit that it can reliably modelled theoretically. \cite{Patton17} performed a Fisher forecast for the one-point convergence PDF based on the L-PICOLA perturbative mocks for a DES-sized survey. They showed that although the degeneracy direction between $\Omega_{\rm m}$ and $\sigma_8$ was very similar to the cosmic shear two-point correlation function, adding information from the PDF improved constraints by about a factor of two compared to two-point statistics alone. Additionally, the presence of measurement systematics was shown to tilt the degeneracy directions between the PDF and the power spectrum. \cite{Liu19WLPDF} focused on the ability of the weak lensing convergence PDF to help constrain the total neutrino mass ($M_\nu$), performing a Markov chain Monte Carlo (MCMC) forecast for a Rubin-Observatory-like survey based on the MassiveNuS simulations. Their analysis included an examination of the benefits of using a range of source redshifts to perform weak lensing tomography. They showed that the PDF alone could provide a 20\% stronger constraint on $M_\nu$ than the power spectrum, and that combining the two measurements could improve the constraint by 35\% over that of the power spectrum alone.

The one-point PDF of the convergence field therefore provides a natural complement to two-point measurements for extracting cosmological information in the mildly non-linear regime. The statistics of weak lensing fields can be difficult to model because of the mixing of different scales (with different degrees of non-linearity) along the line-of-sight. It is well known as an empirical result that the weak lensing PDF can be approximated as lognormal \citep{Taruya02, Hilbert2011}, or by integrating over a lognormally distributed matter field \citep{Xavier2016}. However, to extract cosmological information a more physically principled model of paramaterising the full cosmology dependence of the PDF is required. Having an accurate theory for predicting cosmological statistics presents many advantages. It allows for easier intuitive understanding of the cosmological dependence of the statistic, removes the need for analysis with expensive simulations, and greatly simplifies the performance of cosmological survey forecasts. In this work, we present the first comprehensive forecast for the lensing convergence PDF simultaneously varying all LCDM parameters, as well as individually considering the extensions of a free total neutrino mass $M_\nu$ and dark energy equation of state $w$. Note that beyond galaxy weak lensing, the cosmic microwave background (CMB) lensing PDF \citep{Liu16CMBkappaPDF,Barthelemy_2020CMB} and moments of the thermal Sunyaev-Zeldovich field \citep{Hill13} have also shown promise in cosmological forecasts. Recently, the convergence skew-spectrum has also been computed for the weak lensing of galaxies and the CMB in \cite{Munshi_2020}.

Large-deviation theory \citep[LDT, for a general review see][]{touchette2011} has successfully been applied to many cosmic large-scale structure fields in recent years. \cite{Bernardeau_2016} clarified the applicability of the theory to the cosmological density field, demonstrating its connection to earlier works on the calculation of cumulants and modelling the matter PDF using perturbation theory and spherical collapse dynamics \citep{Valageas2002, Bernardeau14}. In \cite{Uhlemann16}, this approach to calculating the PDF was greatly simplified with an analytical approximation and \cite{Codis16b} demonstrated its potential for constraining cosmology. Recently, the power of the matter PDF to constrain the neutrino mass or primordial non-Gaussianities was explored in \cite{Uhlemann_2020} and \cite{Friedrich_2020}, while \cite{Ivanov19} investigated a renormalisation procedure for nonperturbative effects. Although the matter PDF is not a direct observable, it is the foundation of observable quantities like luminous tracer statistics \citep{Repp_2020} and weak lensing fields, with combined approaches like density-split statistics \citep{Gruen18, Friedrich18,Brouwer18troughs} recently showing particular promise. \cite{Uhlemann18cyl} applied the principles of LDT to densities in cylinders, and pointed out that LDT provides a natural basis for the joint modelling of the matter and weak lensing fields, overcoming a fundamental limitation of the lognormal approximation.

\cite{Barthelemy_2020} presented an LDT-based prediction of the top-hat-filtered weak-lensing convergence PDF on mildly non-linear scales building on early work by \cite{BernardeauValageas00}. The light-cone is divided into thin redshift slices, and the cumulant generating function (CGF) of the projected matter density in the resulting thin cylinders is related to that of the 3D matter density field in each slice and summed over. For this work, we adopt this approach \citep{Barthelemy_2020}, which is compatible with density-split statistics from \cite{Friedrich18}. A nulling method based on the BNT transform \citep{BNT} was implemented to remove contributions from very non-linear small scales to significantly improve the accuracy of the theory. Note that \cite{thiele2020accurate} recently presented an alternative analytic model for the weak lensing convergence PDF using a halo-model-based approach. Their model targets a smaller-scale, more non-Gaussian regime and is not accurate enough for the mildly nonlinear scales considered here, but their Fisher forecast shows promising results for constraining the neutrino mass in a Rubin-Observatory-like survey.

The purpose of the present work is to demonstrate how well fundamental cosmological parameters can be constrained through measurements of the weak lensing PDF. We validate the cosmology dependence of our  model against convergence maps obtained from ray-tracing through the \dustgraincosmo~simulations \citep{Giocoli18} and the MassiveNuS simulations \citep{Liu_2018}. After establishing the accuracy and precision of our model, we use it to perform a Fisher forecast for stage 3 cosmic shear surveys such as Euclid. Our forecast is the first for the convergence PDF to simultaneously vary the full set of $\Lambda$CDM parameters.
We demonstrate that the convergence PDF outperforms the two-point correlation function in constraining cosmological parameters. In addition, we compare the information contained in the full PDF with that provided by a finite number of cumulants of the field. We focus on deploying the theoretical model for weak galaxy lensing at a single source redshift, and leave a tomographic analysis for future work.

Our paper is structured as follows: in Section \ref{sec:theoretical_model}, we describe our theoretical model for the smoothed weak lensing convergence PDF, its sensitivity to cosmology, and how it can be extended to account for massive neutrinos. In Section \ref{sec:validation_with_sims}, we validate our convergence PDF model using weak lensing convergence maps extracted from ray-traced N-body simulations. In Section \ref{sec:fisher_forecast} we perform a Fisher analysis for a Euclid-scale survey to quantify the potential cosmological constraints available from the weak lensing convergence PDF, establishing its complementarity to the two-point correlation function and CMB data. We conclude and summarise potential avenues for future work in Section \ref{sec:conclusion}.

\section{Theoretical Model for the Weak Lensing Convergence PDF}\label{sec:theoretical_model}

In this section, we briefly summarise the key equations required for the calculation of the weak lensing convergence PDF from large deviation theory following the formalism developed in \cite{Barthelemy_2020}. A more complete overview of the derivation of these equations is given in Appendix \ref{app:LDT_moments}. 

The lensing convergence $\kappa$ can be calculated from a projection of the matter density contrast between the source and the observer along the line-of-sight. Using the Born approximation and neglecting lens-lens coupling, it can be written in terms of a redshift integral \citep{kappadef}
\begin{equation}
    \kappa(\bm{\hat{n}}) = \int_0^{z_s} {\rm d}z \, R'(z) \omega(z, z_s) \, \delta(z,R(z)\bm{\hat{n}}),
    \label{eq:def-convergence2}    
\end{equation}
where $R(z)$ is the comoving distance to redshift $z$ (equal to the comoving angular diameter distance $d_A(z)$ in a flat universe, as assumed throughout this work) and $R_s$ is the comoving distance to the source redshift, $z_s$. The weight function $\omega$ is defined as
\begin{equation}
\label{eq:weight}
    \omega(z,z_s) = \frac{3\,\Omega_{\rm m}\,H_0^2}{2\,c^2} \, \frac{d_A(z)\,d_A(z_s,z)}{d_A(z_s)}\,(1+z)\,,
\end{equation}
and the comoving distance is a function of the Hubble parameter
\begin{align}
\label{eq:comdist}
R(z) &= \int_0^z \frac{c}{H(z)} dz\ ,\quad R'(z)=\frac{c}{H(z)}\,,
\end{align}
with the Hubble parameter being sensitive to cosmological parameters through the Friedmann equation. Note that for a flat universe, the lensing weight from Equation~\eqref{eq:weight} combined with $R'(z)$ is independent of the Hubble parameter today.

\subsection{Cumulants and the PDF}
The lensing convergence in a cone with a given opening angle $\theta$, $\kappa_\theta$, for sources at redshift $z_s$, can be viewed as a weighted integral of the density contrast $\delta_{\rm cyl}$ in cylindrical slices of the cone with radii $d_A(z)\theta$ and depth $L$. In the small angle approximation it is possible to assume\footnote{This assumption is valid as long as the scale of the radial variation of the lensing kernel function and the statistical properties of the density field is much larger than the transverse size of the beam. This is a generalisation of the Limber approximation.} that $L$ is much larger than $d_A(z)\theta$. For standard cosmological power spectra\footnote{In this context the cumulants are indeed dominated by modes whose wavelength is smaller or about the transverse size of the cylinder.}, the $n$-order cumulants of the density $k_n^{\text{cyl}}(d_A\theta, L, z)$  in such cylinders will scale like $1/L^{n-1}$. This is true in particular for the cylindrical variance $\sigma^2(d_A\theta, L, z)$, which reads
\begin{equation}
    \sigma^2(d_A\theta, L, z) = \frac{1}{L} \,\int \frac{{\rm d}^2\bm{k}_{{\perp}}}{(2\pi)^2} P(k_{{\perp}},z) \, W_{\rm 2D}^2(d_A\theta k_{\perp}),
    \label{eq:cylindrevariance}
\end{equation}
where $W_{\rm 2D}(l) = 2J_1(l)/l$ is the Fourier transform of the cylindrical top-hat kernel, $J_1$ is the Bessel function of first order, $\bm{k}_{{\perp}}$ are the wave vector modes perpendicular to the line-of-sight and $P(k,z)$ is the matter power spectrum at redshift $z$. 

More generally, the cumulant generating function (CGF) of the lensing convergence, $  \phi_{\kappa,\theta}$, can be straightforwardly related to the CGF of the density in such cylindrical slices, $\phi_{\delta,\textrm{cyl}}$, as \citep{BernardeauValageas00}
\begin{equation}\label{eq:cgf_integral}
    \phi_{\kappa,\theta}(y) = \int^{z_s}_0 {\rm d}z\,R^{\prime}(z)\phi_{\delta,\textrm{cyl}}(\omega(z,z_s)y,d_A(z)\theta,z),
\end{equation}
where $\omega(z,z_s)$ is given in Equation~\eqref{eq:weight} and $\phi_{\delta,\rm cyl}$ is evaluated for cylindrical slices of size $d_A(z)\theta$.
A prescription for the CGF of the density in cylinders is provided by large deviation theory (LDT, for a more detailed outline see Appendix \ref{app:LDT_moments}), which describes the rate function  $\Psi_{\rm cyl}$ (i.e. the leading behavior of the logarithm of the PDF) as some driving parameter (here the variance) goes to zero. Formally, the CGF and the rate function are Legendre transformations of one another.

The determination of the rate function $\Psi$ in the case of a cylindrical density relies on a key property of LDT known as the contraction principle. In this particular context, the contraction principle means that the appropriate rate function for the late-time density field can be computed from the initial conditions using only the most likely mapping between the initial and final densities $\rho=1+\delta$ in cylinders, which we assume to be the cylindrical collapse based on symmetry arguments \citep{Valageas02, Barthelemy_2020}. With this, the rate function for the late-time density in cylinders (assuming Gaussian initial conditions\footnote{Primordial non-Gaussianities can also be straightforwardly accounted for in this formalism as shown by \cite{Uhlemann18pNG,Friedrich_2020}}) takes the form 

\begin{equation}\label{eq:ratefunc}
    \Psi_{\rm cyl}(\rho) = \frac{\tau^2(\rho)}{2\sigma^2_{l}(r_{ini},z)} \,,
\end{equation}
where $\sigma^2_{l}$ is the linear cylindrical variance in the Lagrangian radius $r_{ini}$ (through mass conservation, $r_{ini}=d_A\theta\rho^{1/2}$), and $\tau$ is the most likely linear density contrast corresponding to the final density $\rho$. The relevant equations for cylindrical collapse are summarised in Appendix A of \cite{Friedrich18}; a robust approximation was provided by \cite{Bernardeau95ang}
\begin{equation}\label{eq:cylcollapseparam}
    \zeta(\tau) = \left(1-\frac{\tau}{\nu}\right)^{-\nu} \Leftrightarrow \tau(\rho)=\nu\left(1-\rho^{-\frac{1}{\nu}}\right)\,.
\end{equation}
The value of $\nu$ is set to 1.4, so as to reproduce the value of the tree-order skewness for cylindrical symmetry that is computed from perturbation theory \citep{Uhlemann18cyl}. 
The linear cylindrical variance can be calculated from Equation \eqref{eq:cylindrevariance} when the power spectrum is replaced by the linear power spectrum $P_l(k,z)$, which can be obtained using Boltzmann codes like CLASS \citep{CLASS}.

Although the LDT results become exact only in the limit $\sigma\rightarrow 0$, previous work has shown that the formalism also performs well when extrapolated to non-zero values of the variance. This phenomenological extrapolation, beyond LDT, has to be performed in principle for each random variable -- i.e each redshift slice -- and its corresponding driving parameter, namely the non-linear variance at that redshift. This means that a modelling of the evolving non-linear variance is needed as an external input to our formalism. In practice, a rescaling of the CGF has to be performed to ensure a correct non-linear variance. We calculate the non-linear $\kappa$ variance, $\sigma^2_{nl}$, using Equation \eqref{eq:cylindrevariance} with the nonlinear power spectrum from Halofit \citep{Takahashi_halofit_2012, Bird12}. In \cite{Barthelemy_2020}, the correction for the non-linear variance was performed by rescaling the CGF with the correct projected non-linear $\kappa$ variance, either from Halofit or as measured from simulations. In the present case, we modify this approach slightly and rescale the CGF in each redshift slice by the non-linear variance at the scale $d_A(z)\theta$ 
\begin{equation}
    \begin{aligned} \phi_{\delta,\text{cyl,nl}}(y) &=\!\frac{\sigma^2_{\mathrm{l}}(d_A(z)\theta)}{\sigma^2_{\mathrm{nl}}(d_A(z)\theta)} \phi_{\delta,\text{cyl,l}}\left(\frac{\sigma^2_{\mathrm{nl}}(d_A(z)\theta)}{\sigma^2_{\mathrm{l}}(d_A(z)\theta)} y\right)\!. \end{aligned}
    \label{NLPhi}
\end{equation}

This is the same approach as adopted in \cite{Friedrich18} and implemented in the publicly available code CosMomentum\footnote{\url{https://github.com/OliverFHD/CosMomentum}}, with which we cross-check some of our results. 

Once the CGF has been constructed, the convergence PDF is obtained using an inverse Laplace transform
\begin{equation}\label{eq:laplace_trans}
    P(\kappa) = \int^{+i\infty}_{-i\infty}\frac{\textrm{d}y}{2\pi i}\exp(-y\kappa+\phi_{{\kappa},\theta}(y)).
\end{equation}
The inverse Laplace transform requires an analytical expression for the CGF so as to be able to perform an analytic continuation in the complex plane. In \cite{Barthelemy_2020} this was achieved through the fitting of a double-power-law approximation to the linear variance. In the present work, we instead move to a method first suggested in \cite{BernardeauValageas00}, which involves reframing the derived CGF in terms of an effective mapping $\zeta_{\rm eff}(\tau)$ between an initial unsmoothed field and the final smoothed convergence field \citep[for a detailed description see Appendix \ref{app:effective_mapping} or][]{barthelemy2020}. A polynomial fit of the effective mapping is used to perform the inverse Laplace transform numerically. 

\subsection{Cosmology Dependence}

To get preliminary insights into the cosmology dependence of the PDF derived as above, one can examine the dependence of the cumulants of cylindrical densities and the lensing convergence on the cosmological parameters. In this section, we consider the cold dark matter fraction $\Omega_{\rm cdm}$, the power spectrum amplitude $\sigma_8$ and the dark energy equation of state $w_0$.
Specifically, we focus here on understanding the cosmology dependence of the PDF through the effects of the cosmological parameters on its second and third cumulants (the variance and $\kappa_3$), extending to the dark energy equation of state what was pioneered in \cite{1997A&A...322....1B}.
From Equation \eqref{eq:ratefunc}, it is clear that the density PDF in the context of LDT is intimately related to both the scale-dependence of the linear variance and the dynamics of the cylindrical collapse. More precisely, different densities $\rho$ scan the linear variance at scales $r\rho^{1/2}$ for a range of values around the radius $r$. This is encapsulated in the behaviour of the tree-order perturbation theory prediction for the reduced skewness, $S_3$, of the density at scale $r$. In an EdS universe, this quantity is determined by the first logarithmic derivative of the linear variance \citep{Bernardeau94smoothing,Bernardeau95ang}
\begin{align}
\label{eq:S3pred}
    S_3^{\rm 2D}(r) = \frac{\langle\delta^3(r)\rangle}{\langle\delta^2(r)\rangle^2} =\frac{36}{7} + \frac{3}{2}
    \frac{d\log \sigma^2_l(r)}{d\log r}\,.
\end{align}
In principle, the first term of this expression depends on the expansion rate of the Universe and therefore on the cosmological parameters. This dependence has been found to be very mild in practice \citep{Bernardeau92}, and we have verified that it makes no difference to our results by comparing with output of the CosMomentum code, which implements exact spherical collapse solutions. In the context of weak lensing observations, the main dependence on the expansion rate of the Universe will be encoded in the projection effects.

\begin{figure}
\includegraphics[width=1.\columnwidth]{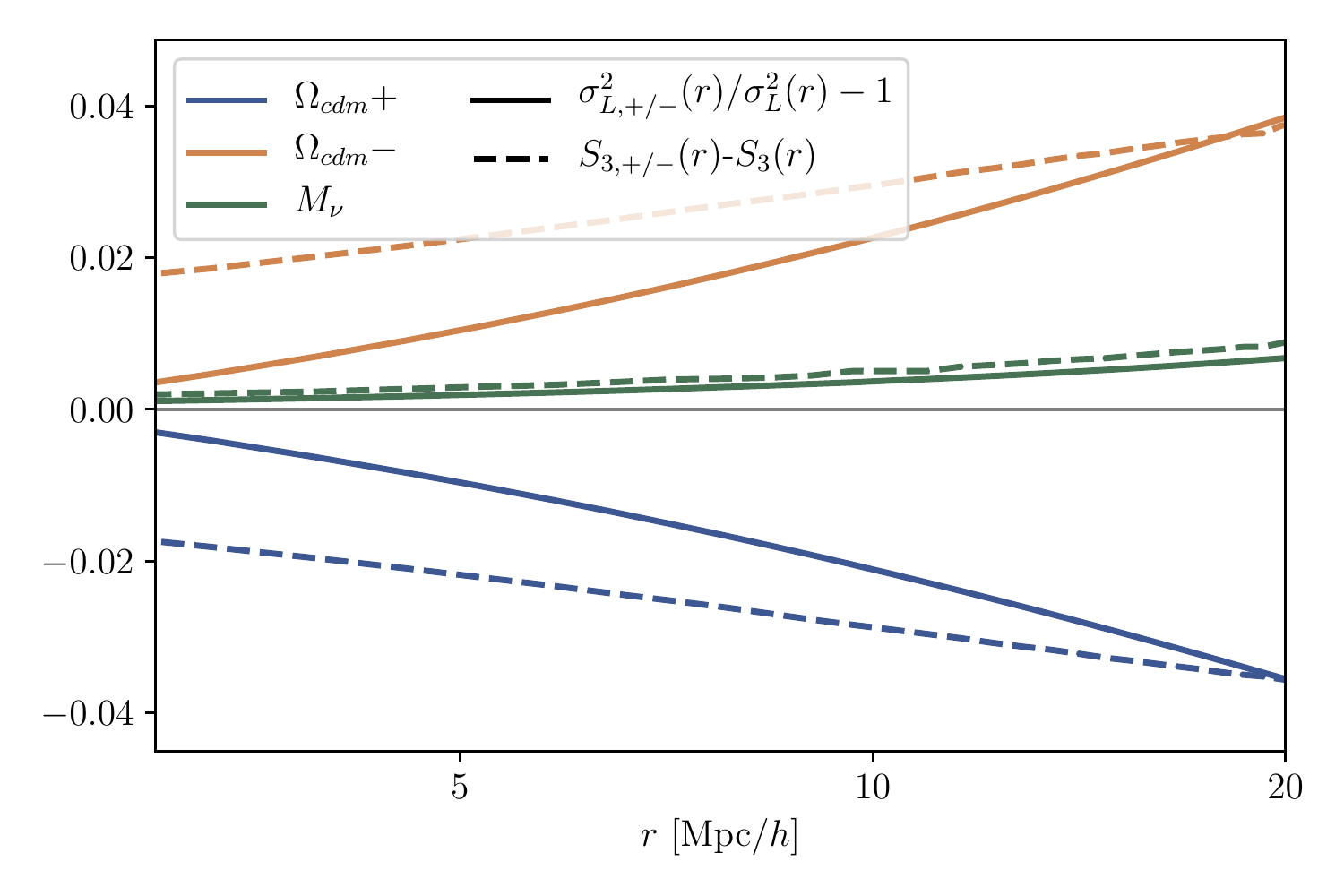}
   \caption{Solid lines: The response of the linear cylindrical variance $\sigma_L^2$ at $z=0$ to different changes in cosmology (denoted by colours) from the fiducial cosmology. The increments used are those specified in Table \ref{tab:cosmology} (using the smaller increment for $\Omega_{\rm cdm}$). Dashed lines: The corresponding differences in the reduced skewness $S_3$ as calculated using Equation \eqref{eq:S3pred}.}
   \label{fig:sigmalincomparison}
\end{figure}

Note that $S_3^{\rm 2D}(r)$ does not depend on the overall amplitude of the density fluctuation $\sigma_8$ as its dependence cancels out in the logarithmic derivative. It depends however on the scale-dependence of the variance. In Figure~\ref{fig:sigmalincomparison}, we show how $\Omega_{\rm cdm}$ impacts the linear variance at $z=0$ and hence the reduced skewness. Note that the shape of the full PDF is also sensitive to a combination of the reduced kurtosis $S_4$ and higher order cumulants, which depend on higher order logarithmic derivatives.

As mentioned before, the main source of dependence on the expansion rate is encoded in the projection effects, as shown in Equation \eqref{eq:cumkappa} at the level of the CGF, or at the level of the cumulants as shown hereafter. In general, the cumulants of $\kappa$ can indeed be related to the cumulants in  line-of-sight cylinders. Viewing the projected density as the sum of densities in such cylinders, it can be shown that
\begin{equation}\label{eq:cumulant_kappa}
    k_n^{\kappa}(\theta,z_s) = \int_0^{z_s}\!\!\!{\rm d}z R'(z)\omega^n(z,z_s)k_n^{\text{cyl}}(d_A(z)\theta, L, z)L^{n-1},
\end{equation}

where $k_n^{\text{cyl}}(d_A(z)\theta,L,z) = \langle \delta_{\text{cyl}}^n (d_A(z)\theta,L,z) \rangle_c$ are the cumulants of the 3D density contrast filtered in a cylinder of transverse size $d_A\theta$ and depth $L$, with a well-defined limit $L\rightarrow \infty$. We can rewrite the cumulants for cylinders in terms of their reduced counterparts, $S_{n}^{\text{cyl}}$ (see also Equation \eqref{eq:redcum}), which are then independent of $L$. At tree order, they are furthermore expected to be redshift-independent and given by their 2D expression (Equation \eqref{eq:S3pred}) so that
\begin{align}
\label{eq:cumkappa}
    k_n^{\text{cyl}}(d_A\theta,L,z)&=S_n^{\text{cyl}}(d_A\theta,z) \sigma^{2(n-1)}_{nl}(d_A\theta,L,z)
 \\
\nonumber &\simeq S_n^{2D}(d_A\theta) \sigma^{2(n-1)}_{l}(d_A\theta,L) D^{2(n-1)}(z).
\end{align}

\begin{figure}

\includegraphics[width=1.\columnwidth]{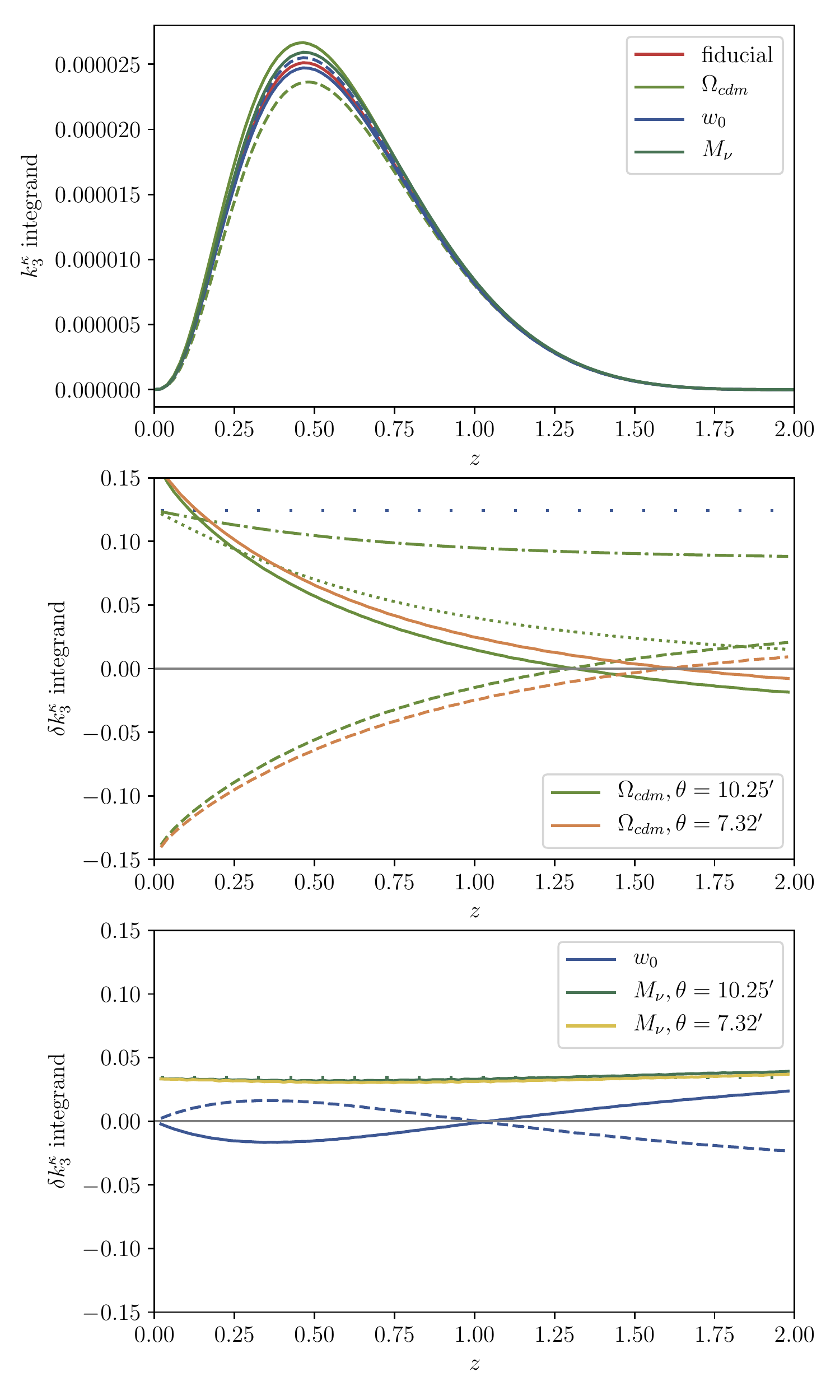}

   \caption{
   (Upper panel) The integrand used to calculate the third cumulant of $\kappa$ in Equation~\eqref{eq:cumkappa}
   for various cosmologies for $\theta=10^\prime$. The solid lines correspond to increases in the corresponding cosmological parameters (by the smaller increments in Table \ref{tab:cosmology}) and the dashed lines to decreases.
   (Middle panel) Fractional changes in the integrand induced by changing $\Omega_{\rm cdm}$ for the two smoothing scales considered for our Fisher forecast. The solid and dashed lines correspond to increases and decreases in the cosmological parameters once again. The additional lines represent the various contributions to the overall differences (see the main text). (Lower panel) Fractional changes induced in the integrand by increasing or decreasing $w_0$, and by changing $M_\nu$ from 0 to 0.15 eV. The changes with $w_0$ are independent of smoothing scale (so only one set of lines are shown) and the changes with $M_\nu$ vary little with smoothing scale (compare the dark green and yellow lines). The $M_\nu$ integrand is almost constant with redshift. The dotted dark green line shows the effect of changing only the $\Omega_{\rm m}$ factor in the integrand.}
   \label{fig:kappa3comparison}
\end{figure}

It is then possible to give the explicit dependence of $k_3^{\kappa}$ on any cosmological parameter. More specifically, the upper panel of Figure \ref{fig:kappa3comparison} shows the integrand for $k_3^{\kappa}$ for various cosmologies. The lensing kernel for a source redshift $z_s=2$ peaks around $z=0.5$. The middle panel shows the fractional changes in the integrand induced by varying $\Omega_{\rm cdm}$. The plot dissects the different contributions for $\Omega_{\rm cdm}$: the contribution from the $\Omega_{\rm m}$ factor in the lensing weight (Equation \eqref{eq:weight}; sparsely dotted line), which is counteracted by the change in the linear growth factor $D(z)$ (see Equation \eqref{eq:cumkappa}; dot-dashed line) and the comoving distance $R(z)$ (see Equation \eqref{eq:comdist}; tightly dotted line). The total change for $\Omega_{\rm cdm}$ depends on the smoothing scale due to the impact of the cylindrical reduced cumulant $S_3$ and the scale-dependent variance $\sigma_{\rm cyl}^2$. The lower panel of Figure \ref{fig:kappa3comparison} shows the fractional changes in the integrand that result from varying $w_0$ and $M_\nu$. In this case, we see that the change induced by $w_0$ is independent of the smoothing scale. This is a shared property of all the moments, which allows us to disentangle the $\Omega_{\rm cdm}$ and $w_0$ contributions from each other even at a single source redshift. Note that while changes in $\Omega_{\rm cdm}$ and $w_0$ lead to a redshift-dependent change in the integrand of the moments, a change in $\sigma_8$ causes a redshift-independent enhancement or diminution. We also emphasise that while we have used the linear variance here for illustrative purposes, our model consistently includes the nonlinear variance.

\subsection{Extending the Model to Include Massive Neutrinos}\label{sec:theoretical_model_neutrinos}

Including massive neutrino cosmologies requires a few adjustments in the implementation of the theory. Massive neutrinos affect the PDF in multiple ways. 

First, the total matter density fraction $\Omega_{\rm m}$ now contains a contribution from $\Omega_\nu$. $\Omega_{\rm m}$ enters into the lensing kernel through Equation \eqref{eq:weight}. This is by far the dominant effect, as can be seen in Figure \ref{fig:kappa3comparison} by comparing the solid dark green line (the overall effect of $M_\nu$ on the $k_3$ integrand) and the sparsely dotted dark green line (the effect of changing $\Omega_{\rm m}$ alone). Because the effect of changing $\Omega_{\rm m}$ dominates, the effect of changing $M_\nu$ in Figure \ref{fig:kappa3comparison} is relatively insensitive to redshift and smoothing scale. This creates a natural degeneracy with $\sigma_8$.  

Next, the scaling of the comoving distance $R(z)$ with redshift changes because the Hubble parameter $H(z)$ changes. The scaling of $H(z)$ is now more complicated, as massive neutrinos contribute with a radiative equation of state at early times and gradually transition into forming a contribution like cold dark matter at late times. For an analytic solution to the evolution of the energy density of neutrinos that can be inserted into the Friedmann equation, see \cite{Slepian_2018}. The simplest solution in this case is to use the Boltzmann code CLASS to extract the appropriate evolution of $H(z)$ and $R(z)$ with redshift.

The unique evolution of massive neutrinos from ultra-relativistic to non-relativistic particles over the course of the history of the Universe has a distinctive effect on the growth of structure. At early times, neutrinos free-stream out of gravitational perturbations and therefore do not contribute to the growth of structure, but they do cluster at late times. This leads to a distinctive scale-dependence in the formation of structures, even on linear scales, with the growth of structure on smaller scales being relatively suppressed. Ultimately, this means that the linear variance $\sigma_l^2(R,z)$ can no longer be evolved with redshift through multiplication by a scale-independent growth factor $D^2(z)$, as the growth factor becomes mildly scale-dependent. For practical implementation, the scale-dependence of the variance must now be independently measured in each redshift slice. 

Finally, massive neutrinos make a small contribution to the cylindrical collapse mechanism (which we parametrise as in Equation \eqref{eq:cylcollapseparam}). \cite{LoVerde14} showed that for neutrino masses within realistic current bounds, the main effect of massive neutrinos on spherical collapse is to increase the collapse threshold by at most 1\%. The impact of such a change can be estimated by varying the parameter $\nu\propto \delta_c$, and \cite{Uhlemann_2020} showed that such a change resulted in a change in the bulk of the 3D matter PDF of less than 1\%. For cylindrical collapse, we assume that the effect is also negligible, and find this approximation sufficient to achieve good accuracy for the bulk of the convergence PDF in the next section.

\section{Validating the Model with Simulations}\label{sec:validation_with_sims}
\begin{table}
\begin{tabular}{l|r|r|r}
\hline
Parameter & Fiducial value & Small Increment & Large Increment\\
\hline
$\Omega_{\textrm{cdm}}$ & 0.26436 & 0.0125 & 0.1\\
$\sigma_8$ & 0.842 & 0.034 & 0.135\\
$w$ & -1 & 0.04 & 0.16 \\
$M_\nu$ (eV) & 0 & 0.15 & - \\ 
\hline 
$h$ & 0.6731 & 0.005 & - \\
$n_s$ & 0.9658 & 0.01 & - \\
$\Omega_{\textrm{b}}$ & 0.0491 & 0.0022 & - \\
\end{tabular}
\caption{The fiducial cosmology and parameter increments chosen for our calculations, largely chosen for consistency with the \dustgraincosmo~simulation suite. Those simulations (see Appendix \ref{sec:howls} for more details) were performed with two step sizes for the three main parameters that were varied ($\Omega_{\textrm{cdm}}$, $\sigma_8$ and $w$) as shown. We do not have simulations that vary the parameters below the horizontal line ($h,n_s,\Omega_{\rm b}$), but calculate their effects theoretically in Section \ref{sec:fisher_forecast} using the step sizes shown.}\label{tab:cosmology}
\end{table}

Before proceeding with a forecast on cosmological parameters, we perform an extensive validation of our theoretical model with simulations. We use sets of weak lensing convergence maps from two suites of ray-traced N-body simulations to determine the accuracy of our theoretical PDFs and their Fisher derivatives: \dustgraincosmo~\citep[see Appendix \ref{sec:howls} and][]{Giocoli18} and MassiveNuS \citep[see Appendix \ref{sec:MassiveNuS} and][]{Liu_2018}. Both of those simulations produce convergence maps of relatively small size (25 \deg$^2$ for \dustgraincosmo~and 12.25 \deg$^2$ for MassiveNuS). To complement the small patch maps, we also make use of a set of full sky lensing maps from \cite{Takahashi17} at a fixed cosmology. 

\subsection{The Fiducial Model PDF}\label{sec:validation_fiducial}

We smooth the \dustgraincosmo~$\kappa$ maps with top-hat filters of size 50 and 70 pixels, corresponding to $7.32'$ and $10.25'$, respectively. For MassiveNuS, the same scales correspond to exactly 25 and approximately 18 pixels. The smoothing is implemented as a direct convolution of the map with the smoothing kernel in the \dustgraincosmo~case, and using a harmonic-space convolution method for MassiveNuS, with the difference being purely practical based on the different storage formats used for the maps. After smoothing, we discard all pixels within a smoothing radius of the edge of a given patch. We measure the histogram of the smoothed $\kappa$ maps in 101 linearly spaced bins for the range $\kappa\in[-0.01,0.01]$. In principle, one might prefer a finer binning, that for example allows to accurately infer the moments from the binned PDF. However, one should consider that when extracting the Fisher information the Kaufman-Hartlap factor given by Equation~\eqref{eq:hartlap} below essentially includes a penalty for too many bins due to the finite number of realisations (here 256-500) from which the covariance is estimated. 

\begin{figure}
\centering
\includegraphics[width=1\columnwidth]{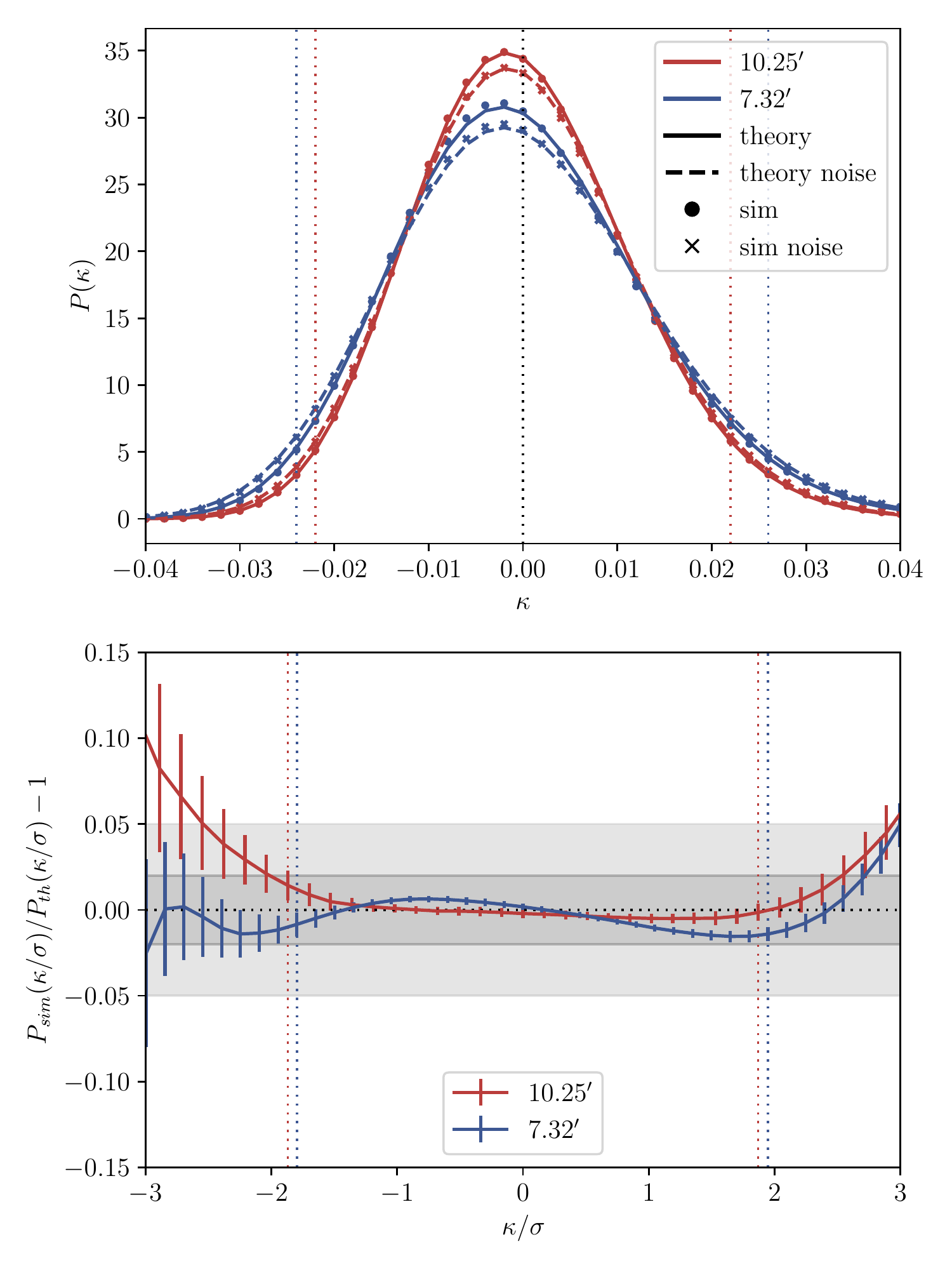}
\caption{(Upper panel) The lensing convergence PDF for the two smoothing scales used in our forecasts for source redshift $z_s=2$ as measured from the mean over 108 realisations of full-sky simulated maps from \protect\cite{Takahashi17}, compared to the corresponding theoretical prediction from large deviation theory (shown both with and without shape noise). (Lower panel) Residuals between the theoretical and simulated noiseless PDFs. The error bars indicate the standard error on the mean computed from the 108 simulations. The vertical coloured lines define the bin ranges we use for our Fisher analysis. The shaded areas in the lower panel highlight 2\% and 5\% ranges.}
   \label{fig:DMPDFfidtheovssim}
\end{figure}

For demonstrating the accuracy of the theory for realistic surveys, it is best to do a comparison of this type with full-sky simulations to avoid additional contributions to the errors from small patch sizes (see Appendix \ref{app:patches}). We therefore compare our theoretical result with that from the simulations of \cite{Takahashi17} in Figure~\ref{fig:DMPDFfidtheovssim}. We show a comparison of the mean PDF over 108 full-sky realisations (long dashed lines) and the theoretical PDF derived from our model using matching cosmological parameters (solid lines) for the two scales 7.32 (blue) and 10.25 (red) arc-minutes. Overall, the prediction and measurement are in very good agreement with the residuals being at a few percent level in the 2-$\sigma$ region around the mean. As expected, the theoretical prediction improves with increasing smoothing scale and the observed departures exhibit a clear signature of a correction to the predicted (tree order) skewness for the smaller scale\footnote{In an Edgeworth expansion of the PDF, the first non-Gaussian correction that appears is proportional to the rms fluctuation times the skewness times an Hermite function of the field of order 3. Any correction to the skewness hence introduces an $H_3$ modulation in the residuals, which is consistent with our measurements.}. 

\begin{figure}
    \centering
    \includegraphics[width=\columnwidth]{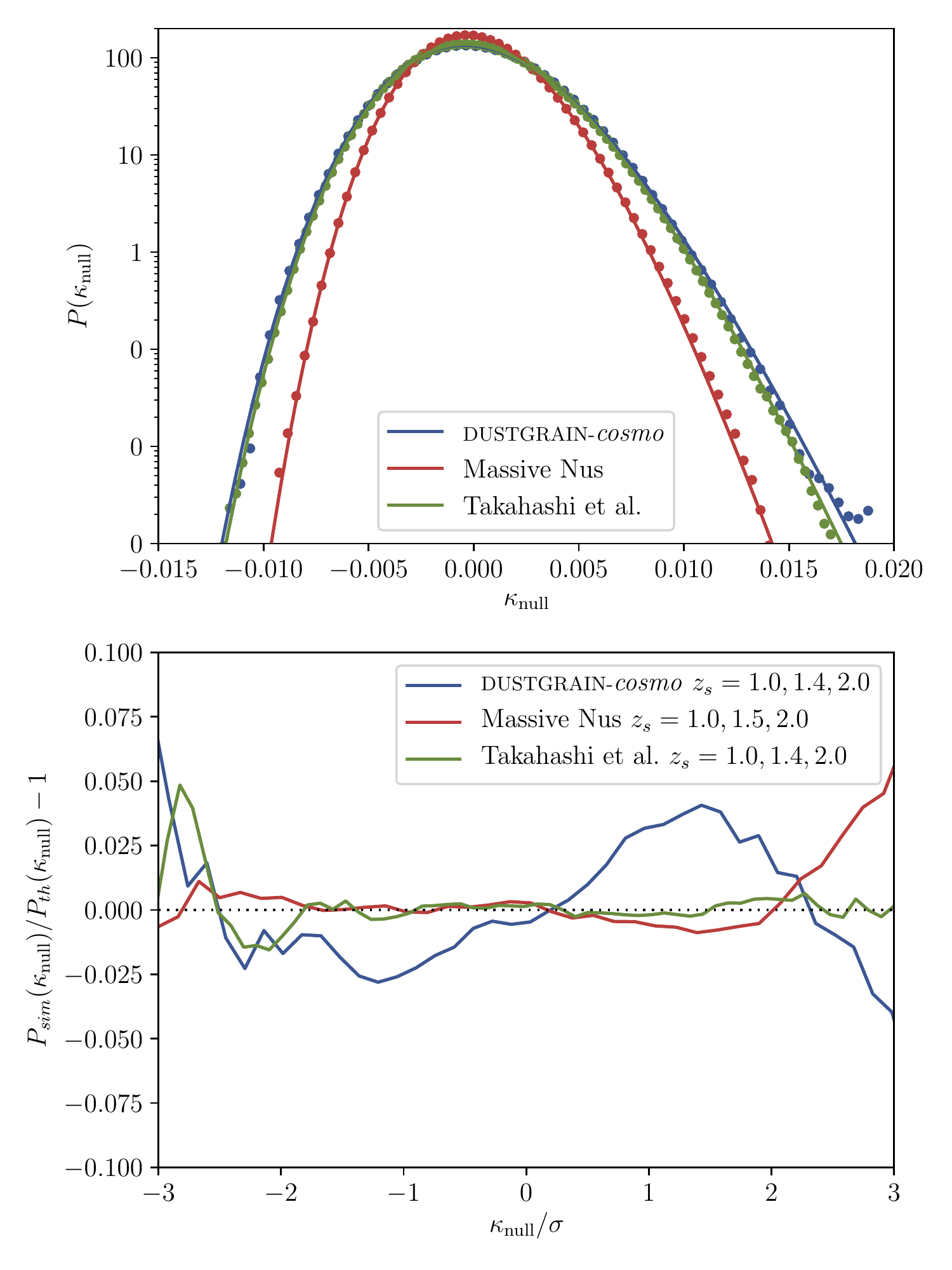}
    \caption{PDFs of the nulled lensing convergence from 256 patches in the \dustgraincosmo~simulations, 1000 patches in the MassiveNuS simulations and 108 full-sky simulations from Takahashi et al. \citep{Takahashi17}. Error bars are neglected because the question of the number of effective independent realisations for the \dustgraincosmo~simulations is unclear (see Appendix \ref{sec:howls} for a discussion of how the simulations are run). It is clear that the simulations of Takahashi et al. and MassiveNuS perform well, but there may be some complications for the \dustgraincosmo~case that fall outside the scope of this work. The nulling source redshifts in each case are given in the captions of the lower panel.}
\label{fig:nullingTakahashi}
\end{figure}

As demonstrated in \cite{Barthelemy_2020}, the mixing of scales involved in integrating the density field along the line-of-sight limits the theoretical accuracy that can be expected. This scale-mixing problem can be eliminated by considering linear combinations of convergence maps at different source redshifts that allow for the `nulling' of small scales and hence facilitate excellent agreement between theory and simulation. The PDF of the nulled convergence $\kappa_{\rm null}$ can be computed using the same formalism as above when replacing the weak lensing weight from Equation~\eqref{eq:weight} by a linear combination of lensing kernels from three different source redshifts $z_s^i$ (called the BNT transform)
\begin{equation}
\omega_{\rm null}(z,\{z_s^i\}) = \sum_{i=1}^3 p_i\, \omega(z,z_{s}^i) \Theta(d_A(z_s^i,z)) \,,   
\end{equation}
where $\Theta$ is the Heaviside step function and the weights $p_i$ are chosen to cancel the contribution from the smallest redshifts and scales by ensuring $\omega_{\rm null}(z<z_s^1)=0$.
We demonstrate this procedure on full-sky simulations from \cite{Takahashi17} in Figure~\ref{fig:nullingTakahashi}, where the maps have been smoothed with the harmonic-space convolution method. As expected, the residual on the bottom panel is very small in this case, below the percent in the 2$\sigma$ range around the mean, despite the PDF being quite non-Gaussian (upper panel). This illustrates the power of LDT to capture the non-Gaussian shape of the convergence PDF with high accuracy in the mildly non-linear regime probed here thanks to the BNT transform.
When applying exactly the same procedure to the maps from \dustgraincosmo
~in Figure~\ref{fig:nullingTakahashi} we find a residual disagreement of a few percent. We suspect that this hints at potential small discrepancies between the different simulations and map-making approaches, whose further study is beyond the scope of this work. As for resolution effects, it is known that the 3D matter PDF at small smoothing scales is impacted at a few percent level by a finite mesh resolution and a finite particle number \citep[see e.g. Figure~5 in][]{Uhlemann_2020}, which are expected to exacerbate the situation for projected lensing quantities that mix small and large scales. See Appendix \ref{sec:sims_and_tools} for details of the simulations and a discussion on the effects of small patch size maps. While \cite{Takahashi17} relies on a full-sky healpix map generated from several runs of nested boxes with varying mass resolution \citep[see Figure~1 in][]{Takahashi17}, \dustgraincosmo~and MassiveNuS maps are created from a single simulation box using randomisation and replication procedures to cover a small lightcone patch. Nonetheless, we see the procedure works well when applied to the convergence maps from MassiveNuS.

Since, observationally, the weak lensing convergence map is obtained from cosmic shear measurements and galaxies themselves are intrinsically elliptical, the observed shear contains a contribution from this intrinsic signal. Shape noise is caused by the variance of the  intrinsic ellipticity, which is the dominant source of noise in shear measurements and impacts the convergence PDF (and its moments and cumulants) as if it was convolved with a Gaussian centred at zero with variance $\sigma_{SN}^2$ \citep{Clerkin16}
\begin{align}
\label{eq:PDFnoise}
    \mathcal P_{\kappa}^{\rm SN}(\kappa) = \frac{1}{\sqrt{2\pi}\sigma_{SN}}\int_{\kappa_{\rm min}}^\infty \!\!\!\! d\kappa' \exp\left(-\frac{(\kappa-\kappa')^2}{2\sigma_{SN}^2}\right) \mathcal P(\kappa')\,.
\end{align}
For the variance describing the shape noise we assume
\begin{equation}
    \label{eq:shapenoise}
    \sigma_{SN}= \frac{\sigma_\epsilon}{\sqrt{n_g \cdot \Omega_\theta}} \propto \theta^{-1}\,,
\end{equation}
where $\sigma_\epsilon=0.30$ is the shape-noise parameter, $n_g$ the galaxy density (here assumed to be $30$ arcmin$^{-2}$) and $\Omega_\theta$ the solid angle (in arcmin$^{2}$). For the simulated maps, we include shape noise by adding a white noise Gaussian random map to the `raw' simulated convergence map before smoothing, keeping the seed fixed for a given realisation when varying cosmologies. For the theoretical predictions, we convolve the theoretical `raw' convergence PDF at the desired smoothing scale with a Gaussian of the appropriate width. Since the area scales quadratically with the radius, the variance due to shape noise is inversely proportional to the smoothing scale. The standard deviation of the noiseless smoothed convergence scales approximately as $\sigma_\kappa\propto\theta^{-1/3}$ for a smoothing scale $\theta$. For a source redshift of $z_s=2$ and smoothing scales of $\theta_1=7.32'$ and $\theta_2=10.25'$, the cosmological signal dominates over the shape noise contribution. Note that due to the symmetry of this convolution, the third moment $\kappa_3$ is only very mildly affected.

In Figure~\ref{fig:DMPDFfidtheovssim} we also demonstrate the impact of shape noise on the smoothed weak lensing convergence PDF for the scales of interest here. As expected, the shape noise broadens the PDF compared to the noiseless case. Note that the addition of shape noise simultaneously decreases the PDF values around the peak (signal) and the PDF covariance (error).

\subsection{The Cosmology-Dependence of the PDF}\label{sec:validation_derivatives}

\begin{figure}
\centering
  \includegraphics[width=1\columnwidth]{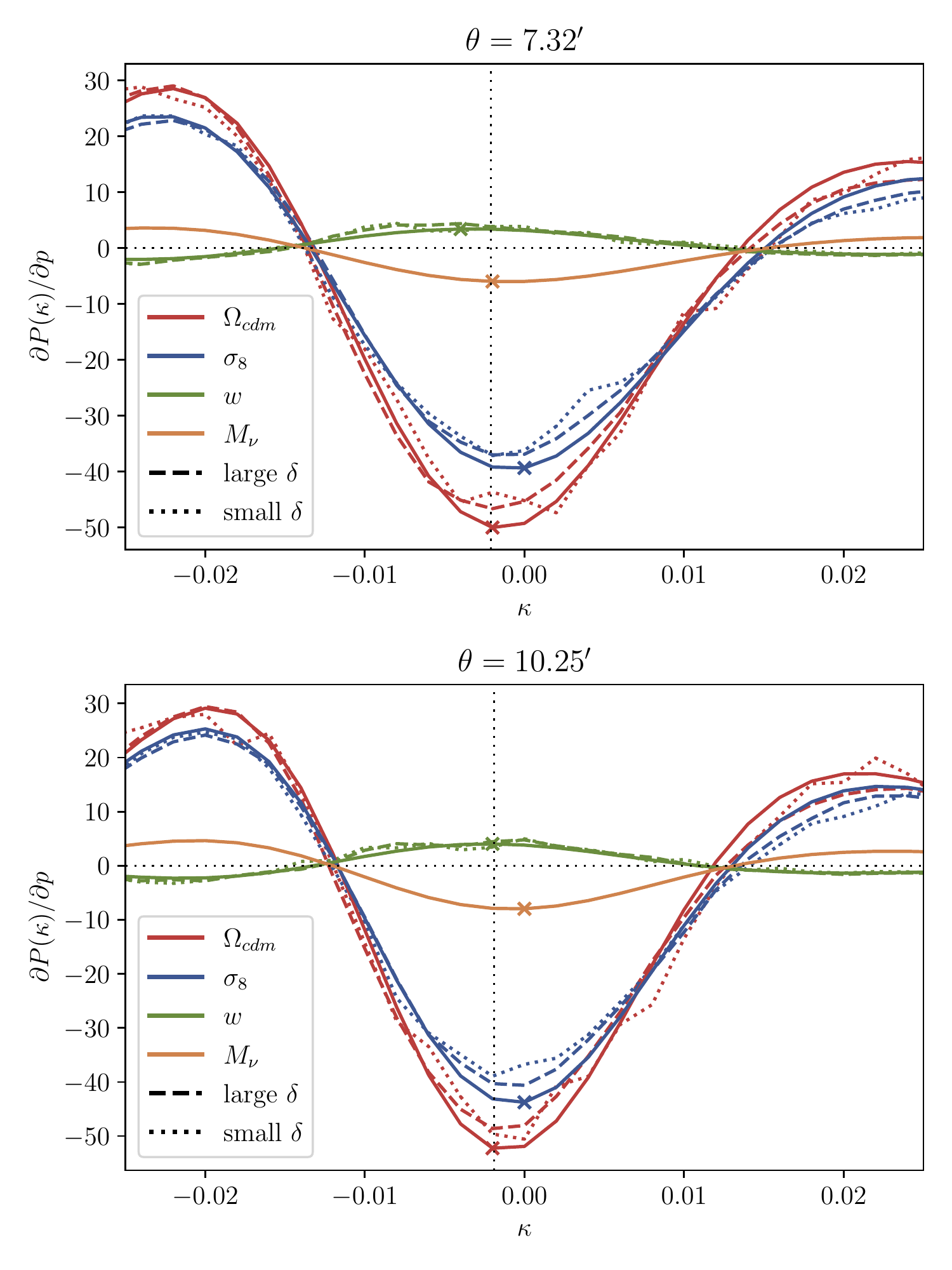}
   \caption{Derivatives of the lensing convergence PDF with respect to $\Omega_{\rm cdm}$, $\sigma_8$, $w_0$ and $M_\nu$ computed for large/small steps in the simulation (dashed/dotted) and the theory for large steps (solid) at two smoothing scales, $\theta=7.32'$ and $\theta=10.25'$. In the $M_\nu$ case, only a theoretical line is shown, and this is validated using the MassiveNuS simulations instead in Figure \ref{fig:PDFderivativeMnu}. One can  see that the small parameter steps in the simulation suffer from numerical noise, which is why we rely on the large steps to validate our theoretical predictions. When calculating the theory derivatives for validation purposes, we include a large-scale cut in $l$ when calculating the variance at $l=36$, corresponding to the patch size of \dustgraincosmo. The vertical dotted lines correspond to the locations of the peaks in the fiducial PDFs. The coloured crosses indicate the extrema of the derivatives. 
   }
  \label{fig:PDFderivatives}
\end{figure}

\begin{figure}
\centering
\includegraphics[width=\columnwidth]{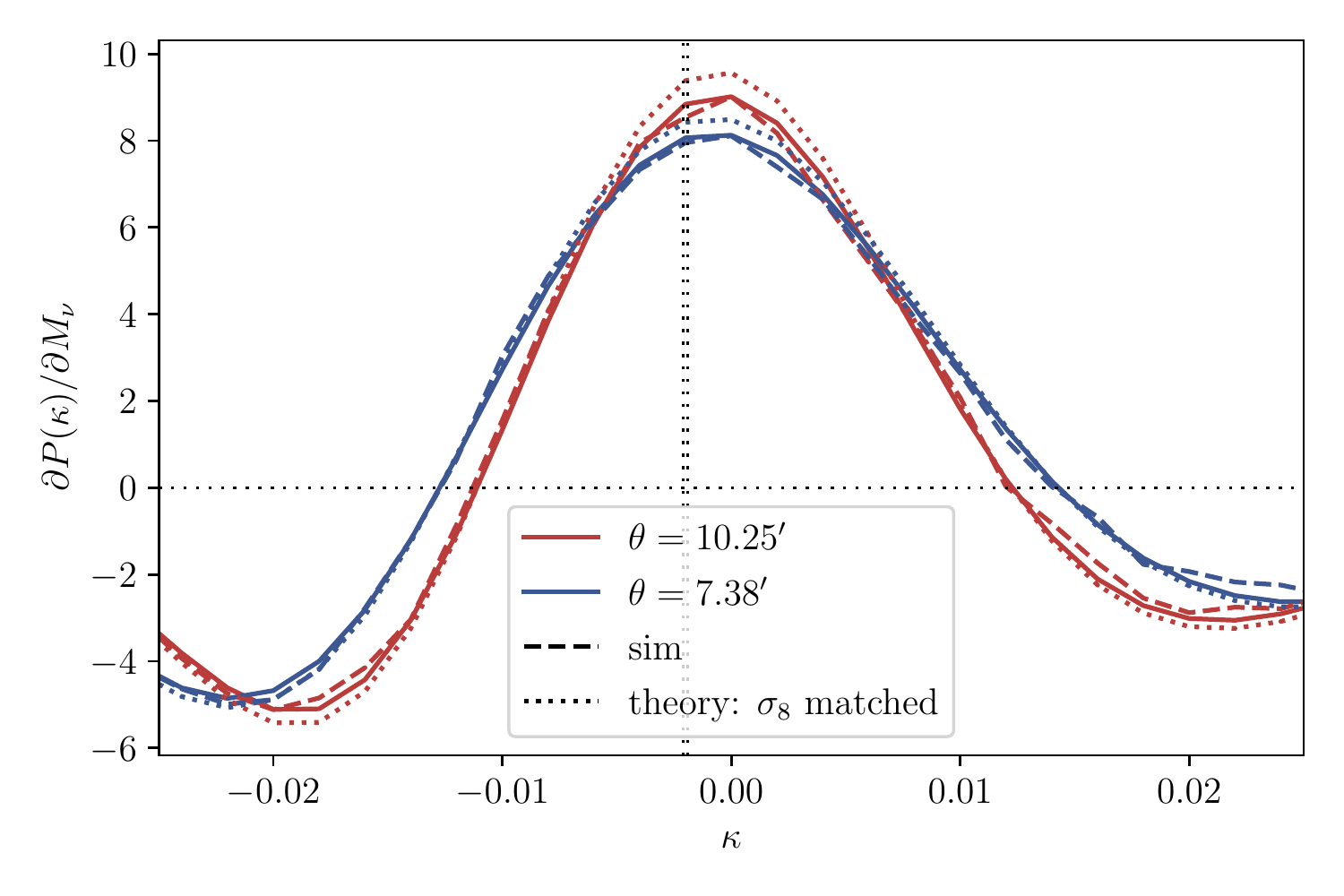}
   \caption{Derivative of the lensing convergence PDF with respect to $M_\nu$ computed as a single-sided derivative using the difference between the fiducial cosmology from MassiveNuS (which has a total neutrino mass $M_\nu$ of 0.1 eV) and their corresponding cosmology without massive neutrinos. This particular comparison is shown only for validation purposes. We calculate the actual theoretical derivatives used in the Fisher matrix calculations in Section \ref{sec:fisher_forecast} for the cosmology given in Table \ref{tab:cosmology}. Here we include a cut in $l$ at $l=60$, corresponding to the MassiveNuS patch sizes. The smaller smoothing scale here is set to 7.38$^\prime$ instead of 7.34$^\prime$ simply because it corresponds to a whole number of pixels for the MassiveNuS resolution. The direction of the derivative is reversed compared to that in Figure \ref{fig:PDFderivatives} because $\sigma_8$ is not fixed here and adding massive neutrinos lowers $\sigma_8$. The dotted \lq$\sigma_8$ matched\rq~lines show how the derivative looks if the PDF with massive neutrinos in the theoretical derivative is replaced by a PDF for a cosmology without massive neutrinos but with equal $\sigma_8$ \citep[$\sigma_8=0.8295$ from][]{Liu_2018}.  The difference between the solid and dotted lines therefore corresponds to the orange line in Figure \ref{fig:PDFderivatives}. The vertical dotted lines correspond to the locations of the peaks in the fiducial PDFs (the leftmost is for the smaller scale).}
  \label{fig:PDFderivativeMnu}
\end{figure}

In this section, we compare the predicted response of the PDF to cosmological parameter changes with measurements from simulations. The PDF response enters the Fisher forecast in Section~\ref{sec:fisher_forecast} in terms of derivatives with respect to cosmological parameters. The calculation of these theoretical derivatives requires several ingredients. For each cosmology (see Table \ref{tab:cosmology}), we generate linear and non-linear (Halofit) matter power spectra from CLASS \citep{CLASS} at a range of redshifts between $z=0$ and our source redshift $z_s=2$. We use these to calculate the linear and nonlinear variance values as a function of redshift following Equation \eqref{eq:cylindrevariance}. CLASS is also used to output $H(z)$ and $R(z)$ for the relevant redshifts. The density CGF is calculated for redshift slices between $z=0$ and $z=2$ in steps of $0.05$ (see Appendix \ref{app:LDT_moments} for details of the calculation of $\phi_{\delta,{\rm cyl}}$ from LDT). These CGFs are then rescaled with the nonlinear variance as outlined in Equation \eqref{NLPhi}, before being combined as in Equation \eqref{eq:cgf_integral} and undergoing an inverted Laplace transform to produce the PDF.

The simulations with we compare our results in this section have produced very small-area maps (5 deg on a side for \dustgraincosmo~and 3.5 deg on a side for MassiveNuS). The variance with which we rescale our CGF must take this into account. In these cases, instead of rescaling by $\sigma_{\rm nl}^2(d_A(z)\theta)$ at each redshift, we rescale with a projected variance derived from Halofit convergence angular power spectra. This allows us to straightforwardly apply a cut in $l$ to exclude the large modes beyond the size of the maps with which we compare. This slightly rougher method is sufficient to validate our derivatives in this section, as we will see. In our Fisher matrix calculations, no cut in $l$ is implemented and the rescaling with the Halofit variance is carried out in each redshift slice individually, but we have validated that whether we rescale with the projected variance or the individual variance in each redshift slice makes no difference to the results, although the latter approach gives slightly better residuals (see Figure \ref{fig:DMPDFfidtheovssim}).

In Figure \ref{fig:PDFderivatives}, we compare the derivatives of the PDF with respect to $\Omega_{\rm cdm}$, $\sigma_8$ and $w_0$ as obtained using the theory and the \dustgraincosmo~simulations at the two relevant scales. We also show a theoretical derivative for $M_\nu$, although we cannot directly validate it with \dustgraincosmo~(we validate it by comparison with MassiveNuS instead in Figure \ref{fig:PDFderivativeMnu}). We find the results to be in very good agreement. We note that with the number of realisations we have, the smaller step sizes provided for the simulations (see Table \ref{tab:cosmology}) result in significant numerical noise (see the dotted lines in Figure \ref{fig:PDFderivatives}). This kind of noise can result in exaggerated constraints by artificially breaking degeneracies between parameters. The larger step sizes provide much smoother derivatives, but we find that even those contain some small amount of noise that can artificially enhance their constraining power by a significant amount when inserted into a Fisher matrix. This presents an obvious advantage to having a theory with which to model observables in such calculations. In addition, our validated complete cosmological model for the PDF allows us to extend our forecasts to parameters beyond those varied in particular simulations (see Section \ref{sec:extended_constraints}). Note that while the general shape of derivatives might look similar at a first glance, there are two hints for potential degeneracy breaking. If derivatives have either different zero crossings or different extrema locations, this means they are not proportional to each other. The first can be seen for example in the positive $\kappa$ region for the the red $\Omega_{\rm cdm}$ and the blue $\sigma_8$ line. The second can be observed from the green line for $w$ whose peak is at a different location from the rest of the dips.

In Figure \ref{fig:PDFderivativeMnu}, we use the MassiveNuS simulations as a basis for determining how well our theory captures the effects of massive neutrinos on the PDF. The fiducial MassiveNuS cosmology \citep[referred to as model 1b in][]{Liu_2018} has cosmological parameters $\Omega_{\rm m}=0.3$, $\sigma_8=0.8295$ and $M_\nu=0.1$ eV. This is used with a model with the same $\Omega_{\rm m}$ but without massive neutrinos, which has $\sigma_8=0.8523$ (model 1a) to calculate the difference shown in Figure \ref{fig:PDFderivativeMnu}. All relevant cosmological parameters are provided in Table 1 of \cite{Liu_2018} and its caption. We see from Figure \ref{fig:PDFderivativeMnu} that the effects of adding massive neutrinos to the cosmology are well captured by the theory. 

\subsection{The Two-Point Correlation Function}\label{sec:validation_2pcf}

We measure the two-point correlation function from the unsmoothed kappa maps using the TreeCorr Python package \citep{TreeCorr}. We use 50 logarithmically spaced bins from 0.1 arcmin to 400 arcmin. To replicate the small patch areas of our two sets of maps, we cut low multipoles $l<l_{\rm cut}=\pi/\theta_{\rm sim}$. For our analysis, we discard scales below $\theta_{\rm min}=5$ arcmin to mitigate small scale inaccuracies and artefacts resulting from our cut of $l_{\rm max}=4096$ when generating FLASK maps Appendix~\ref{sec:flask}. Note that this restriction of scales is similar to the one adopted in \cite{Patton17}, where they performed an $l$-cut to equate the number of independent modes in the PDF and the two-point correlation function. In their case, the PDF smoothing scale of $\theta \simeq 14$ arcmin ($N_{\rm side}=256$) related to $l_{\rm max}=886$ and a corresponding $\theta_{\rm min}\simeq \pi/l_{\rm max}\simeq  12$ arcmin. The cosmological derivatives for the two-point correlation function match those generated using Halofit very well and we use the Halofit derivatives as the input for our Fisher forecasts to avoid simulation noise and ensure compatibility between the PDF and two-point correlation function. 

\section{Fisher Forecast for a Euclid-like Survey}\label{sec:fisher_forecast}

In this section, we quantify the information content of the lensing convergence PDF on some key $w$CDM cosmological parameters using the Fisher matrix formalism. After having validated our theoretical predictions for the PDF and the Halofit two-point correlation function against simulation measurements, we perform our forecast using purely those inputs. This is important because, as mentioned in Section \ref{sec:validation_derivatives}, we find that apparently negligible amounts of numerical noise in the PDF derivatives extracted from simulations can lead to artificially strong constraints. 

The Fisher matrix for a set of cosmological parameters, $\bm{\theta}$, given a (combination of) statistics $\bm{s}$ is defined as
\begin{equation}
F_{ij}= \sum_{\alpha,\beta}\frac{\partial s_\alpha}{\partial \theta_i}C^{-1}_{\alpha \beta}\frac{\partial s_\beta}{\partial \theta_j}~,
\label{eq:Fisher}
\end{equation}
where $s_i$ is element $i$ of the statistic $\bm{s}$ and $C_{\alpha,\beta}$ are the elements of the covariance matrix of measurements $\bm{\hat s}$ of $\bm{s}$, defined as
\begin{equation}
\label{eq:covariance}
C_{\alpha \beta} = \langle (\hat s_\alpha-\langle \hat s_\alpha \rangle)(\hat s_\beta - \langle \hat s_\alpha \rangle) \rangle\ .
\end{equation}
where $\langle\cdot\rangle$ denotes an ensemble average.

The Fisher matrix allows us to determine the error contours on a set of cosmological parameters under the assumption that the likelihood is Gaussian. The inverse of the Fisher matrix gives the parameter covariance. The error on the parameter $\theta_i$, marginalised over all other parameters, is given by
\begin{equation}
\delta\theta_i\geq \sqrt{\left(F^{-1}\right)_{ii}} \ .
\end{equation}

\begin{figure}
    \centering
    \includegraphics[width=1\columnwidth]{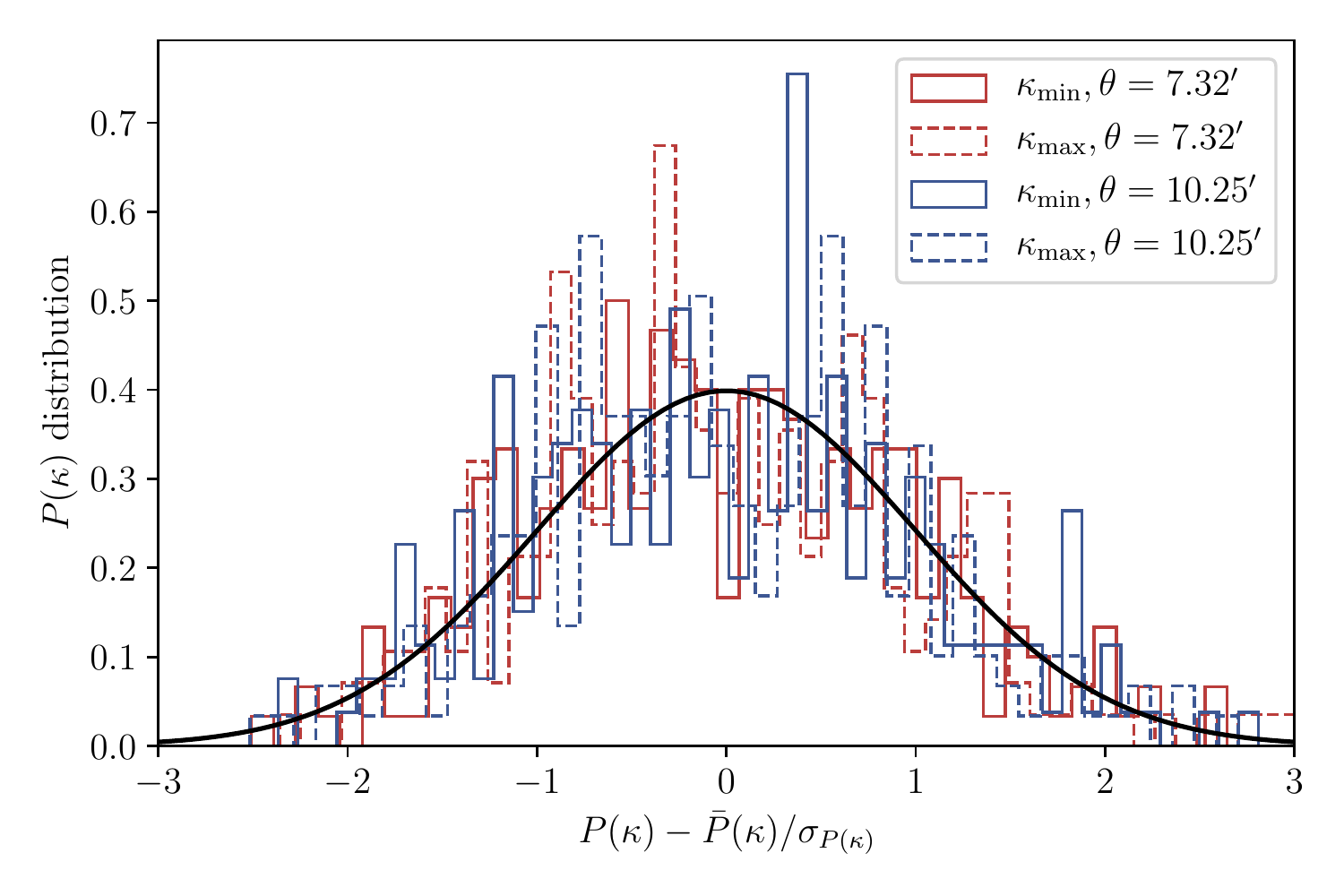}
    \caption{The distribution of realisations around the mean PDF in the lowest (red) and highest (blue) bins for the smaller smoothing scale (solid) and the larger scale (dashed). The black line shows a Gaussian of zero mean and unit variance, which is the expectation for Gaussian noise and a good approximation to all distributions.
    }
    \label{fig:distributionPDFbins}
\end{figure}

A Fisher analysis has a number of limitations: it only yields realistic error bars if measurements of the considered data vectors have Gaussian noise and if the responses of these data vectors to changing cosmological parameters are close to linear.
We demonstrate that the distribution of individual bins of the PDFs measured in the simulations are sufficiently close to a Gaussian distribution in Figure~\ref{fig:distributionPDFbins}. We focus on the lowest and highest two $\kappa$ bins we consider for the Fisher analysis (see the following section) and make a histogram of the fluctuations of the individual realisations around the mean in units of the measured variance. From this figure, we can expect a small impact of residual non-Gaussianity on the total width of our forecasted contours. Note that, in contrast, the moments of the lensing convergence can have strongly non-Gaussian distributions around the mean, in particular in small simulation patches as demonstrated in our discussion of moments-based constraints in Appendix \ref{app:LDT_moments}. 

Fully realistic data analyses might have to account for systematic effects by marginalising over additional nuisance parameters. However, the main focus of this study is to explore the complementarity between the two-point correlation function and the one-point lensing convergence PDF as cosmological probes. Hence, we expect these limitations to have limited impact on our findings.

\subsection{Covariance Matrices}\label{sec:fisher_covariances}

In the following, the data vector $\bm{s}$ entering Equation~\eqref{eq:Fisher} will either consist of histogram bins of measurements of the convergence PDF, of measurements of the two-point correlation function of convergence in a set of angular bins or of a combination of both. In the cases where $\bm{s}$ includes PDF measurements we estimate the covariance matrix that enters Equation~\eqref{eq:Fisher} from a set of $N_{\mathrm{sim}}$ simulated measurements $\bm{\hat s}_i$, $i = 1,\ \dots\ ,\ N_{\mathrm{sim}}$ , as
\begin{equation}
    \hat C_{\alpha,\beta} = \frac{1}{N_{\mathrm{sim}} - 1} \sum_i (\hat s_{i,\alpha} - \bar s_\alpha)(\hat s_{i,\beta} - \bar s_\beta)\ ,
\end{equation}
where $\bm{\bar s}=\sum_i \bm{\hat s}_i/N_{\mathrm{sim}}$ is the mean of all measurements. To compute the Fisher matrix with Equation~\eqref{eq:Fisher}, we need to estimate the inverse covariance matrix or precision matrix. Since matrix inversion is a non-linear operation, the noise in the above estimate of the covariance elements $\hat C_{\alpha,\beta}$ will lead to a bias in the elements of the precision matrix. Through a miracle of nature this bias is just a factor multiplying the entire matrix, the so called 
Kaufman-Hartlap factor \citep{Kaufman67,Hartlap06},
\begin{equation}
\label{eq:hartlap}
    h=(N_{\rm sim} - 2 - N_s)/(N_{\rm sim} - 1) \ ,
\end{equation}
where $N_s$ is the number of data points in $\bm{s}$. Please note that the Kaufman-Hartlap factor \citep[but also the more advanced treatment of][]{Sellentin_2016} only correct the width of parameter contours for bias from covariance estimation noise. This is sufficient for a Fisher analysis like the one presented here, but it is not sufficient for parameter estimation based on an actual measurement of $\bm{s}$. In the latter case, covariance noise will not only impact the width of parameter contours but also introduce additional scatter to the contour location. This needs to be taken into account independently of the Kaufman-Hartlap factor, as e.g.\ described in \cite{Dodelson_2013} or even more completely in \cite{Percival_2014} \citep[see also figure 1 of][for a visualisation of this effect]{Friedrich_2018}.

\begin{centering}
\begin{figure}
\centering
  \includegraphics[width=1.\columnwidth]{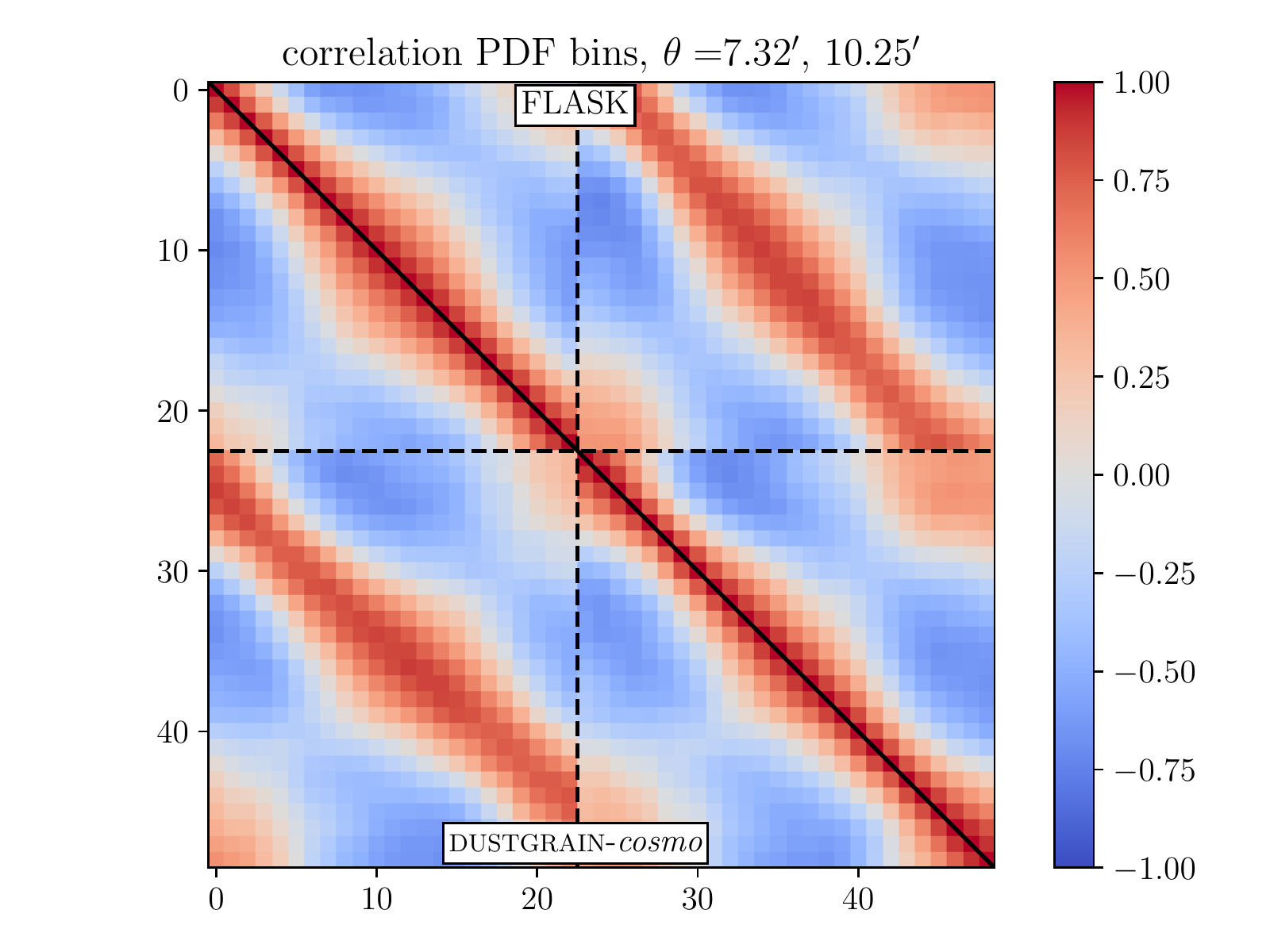}

   \caption{(Upper panel) The reduced cross-correlation matrix (Equation \ref{eq:reducedCMat}) of the weak lensing convergence PDF at radii $\theta_1=7.32'$ and $\theta_2=10.25'$ and source redshift $z_s=2$ without shape noise. We display bins [-0.024,0.026] for $\theta_1$ and [-0.022,0.022] for $\theta_2$, which corresponds to cutting $[2\%,5\%]$ probability in low/high $\kappa$. The black line indicates the diagonal. The lower triangle shows the result from the 256 $\kappa$ maps created from one \dustgraincosmo~N-body simulation with randomisation and the upper triangle shows the result from 500 independent full-sky maps created from FLASK.}
  \label{fig:crosscorrelation_matrix}
\end{figure}
\end{centering}

In Figure \ref{fig:crosscorrelation_matrix} we show the correlation matrix
\begin{equation}
\label{eq:reducedCMat}
\text{Corr}_{ij} = \frac{C_{ij}}{\sqrt{C_{ii}C_{jj}}}\,,
\end{equation}
for measurements of the convergence PDF at our two smoothing scales. This matrix visualizes the amount of correlation throughout the data vector elements.  We compare the correlation matrices obtained from 256 maps from the \dustgraincosmo~simulations and from 500 lognormal maps generated using the publicly available FLASK tool \citep{Xavier2016} in Figure \ref{fig:crosscorrelation_matrix}. In practice, we implement the latter in our Fisher matrix calculations, but we see that both methods are in reasonable agreement. Naturally, in both cases, the measured covariances are re-scaled to the area of a Euclid-sized survey, assuming that the covariance is proportional to survey area. In the two blocks on the diagonal in Figure \ref{fig:crosscorrelation_matrix} we see that, as expected, neighbouring PDF bins are positively correlated, while intermediate underdense and overdense bins are anti-correlated with each other. Note that the tails of the PDF, which are excluded in the plot, are strongly correlated with each other and anti-correlated with the peak.
Additionally, we observe that PDFs at different scales are strongly but not perfectly correlated with each other. This is expected, as the two smoothing apertures partially (but not perfectly) overlap. The cross-correlations between PDF bins of different scales look very similar to the bin correlations for the individual PDFs, because the matter clustering changes mildly with radius \citep[see Figure 5 in][]{Uhlemann17Kaiser}. 

In all contours that only consider the convergence two-point function we use an analytic covariance model in order to circumvent noise (and potential small-scale numerical artefacts) associated with covariance estimation from FLASK. Our covariance model is based on the lognormal covariance model by \citet{Hilbert2011} but uses the procedure of \citet{Friedrich18} to fix the parameters of the lognormal distribution as well as the analytic treatment of sky curvature and bin averaging that is detailed in \citet{Friedrich_in_prep}. We find that our analytic covariance agrees well with the corresponding estimate from FLASK for the range of scales  (bigger than a few arcmin) for which FLASK can be reasonably applied, see Figure~\ref{fig:2pcf_sims_halofit_covariance}.

\subsection{Forecasted Constraints for $\Omega_{\rm m}$, $\sigma_8$, $w_0$ and $M_\nu$}\label{sec:fisher_constraints}

\begin{centering}
\begin{figure}
\centering
  \includegraphics[width=1\columnwidth]{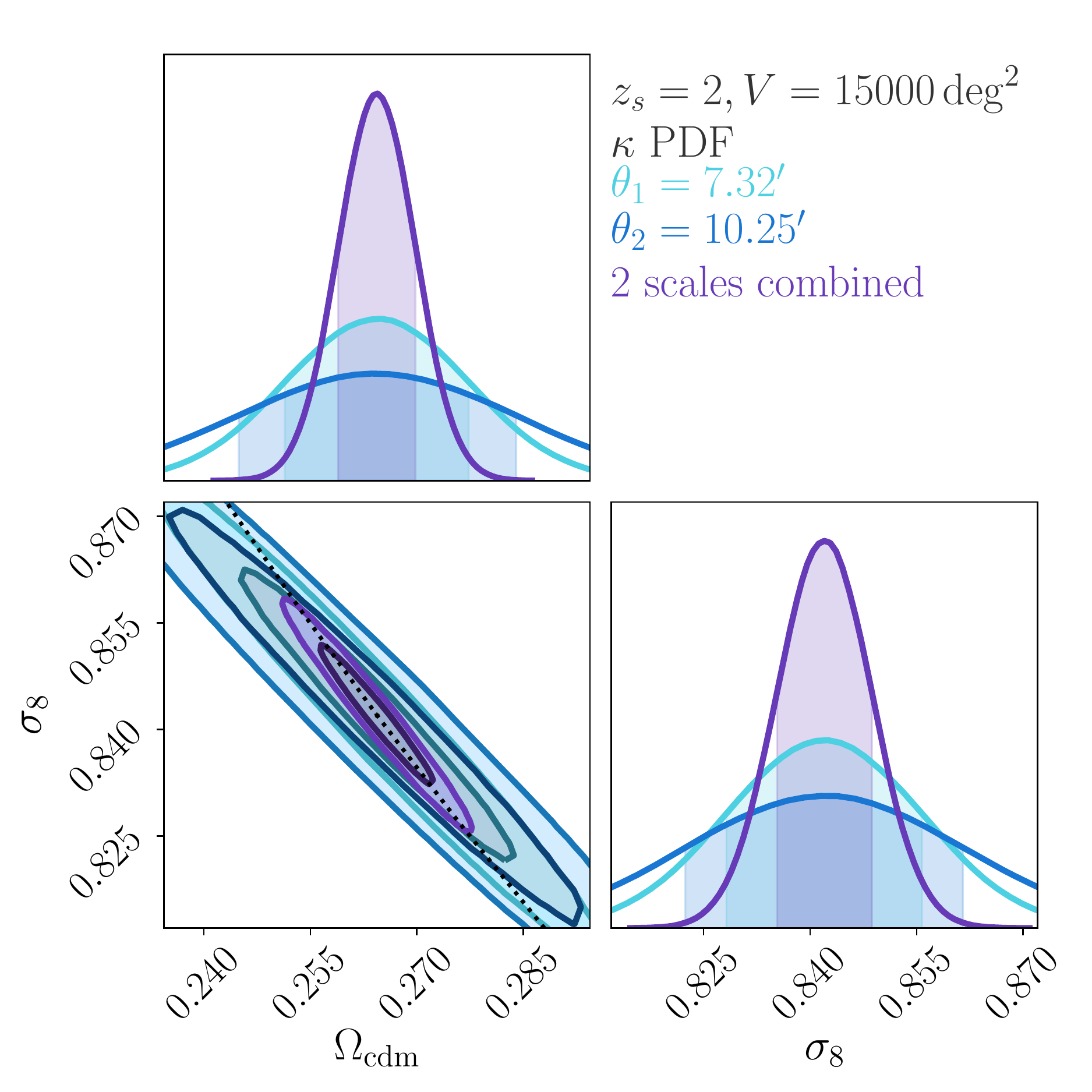}
   \caption{Fisher forecast constraints on $\Omega_{\rm cdm}$ and $\sigma_8$ for a Euclid-like survey from the weak lensing convergence PDF at $z_s=2$. Contours are shown for the PDF measured for a single smoothing scale of $\theta_1=7.32^\prime$ (cyan), a single scale of $\theta_2=10.25^\prime$ (blue), and the two scales combined (purple). It is clear that the use of two smoothing scales significantly improves the constraints. The combined PDF contour is oriented along the line of constant $\Sigma_8=\sigma_8\sqrt{\Omega_{\rm m}}$ (black dotted line).
   }
  \label{fig:contour_omega_cdm_sigma8_scales}
\end{figure}
\end{centering}

We now analyse the cosmology-constraining power of the weak lensing convergence PDF and its complementarity to the two-point correlation function. Figure \ref{fig:contour_omega_cdm_sigma8_scales} shows constraints from the convergence PDF for two individual smoothing scales and their combination. It is clear that combining two smoothing scales helps reduce the degeneracy between $\Omega_{\rm cdm}$ and $\sigma_8$ because the $\Omega_{\rm cdm}$ derivative is a function of the smoothing scale (see Figure \ref{fig:kappa3comparison}). We see that the degeneracy direction aligns roughly with the $\Sigma_8=\sigma_8\sqrt{\Omega_{\rm m}}$ direction. 

\begin{centering}
\begin{figure}
\centering
  \includegraphics[width=1\columnwidth]{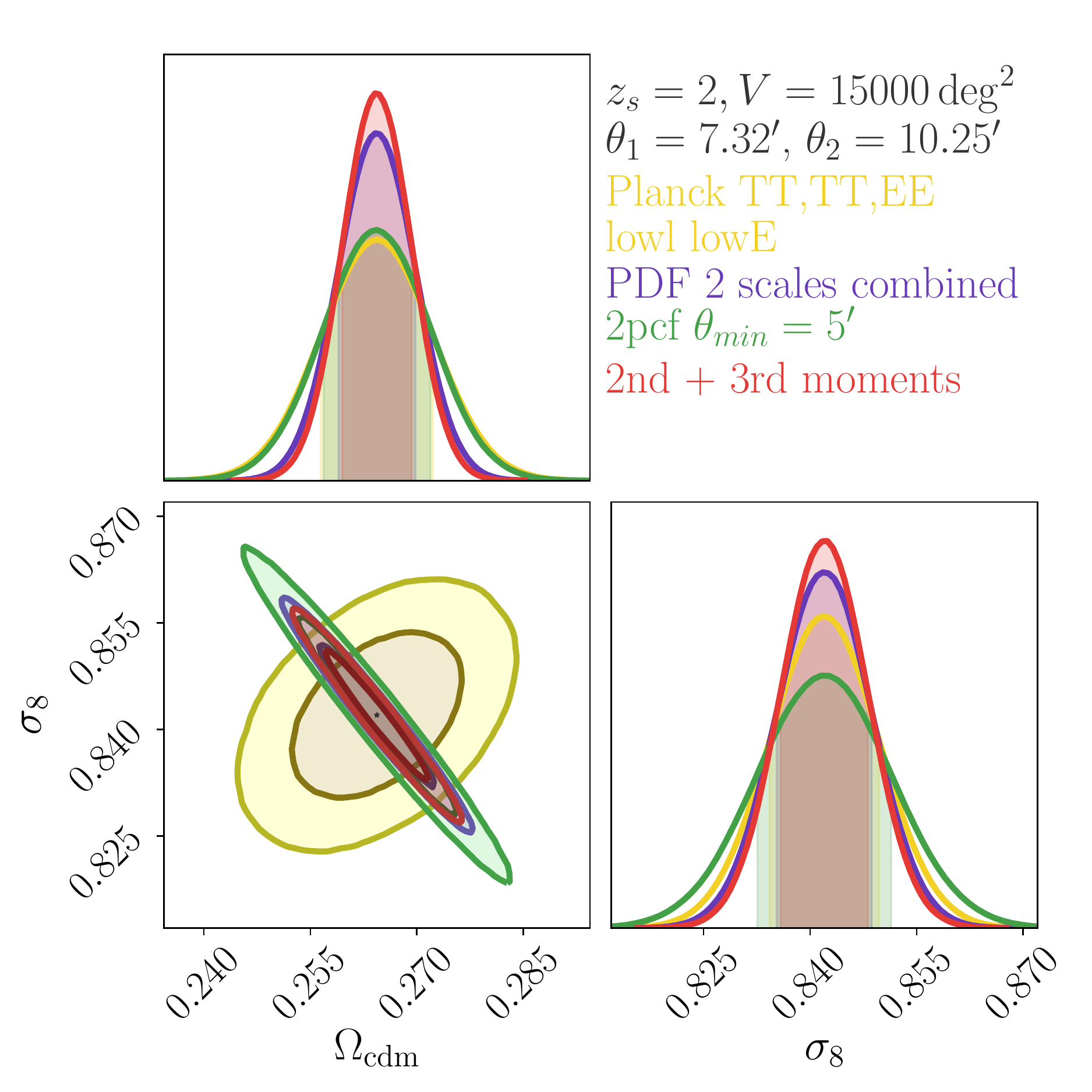}
   \caption{As Figure \ref{fig:contour_omega_cdm_sigma8_scales}, but presenting a comparison of the constraints from the PDF with two smoothing scales (purple) to constraints from other probes. Contours for the two-point correlation function with $\theta_{min}=5^{\prime}$ are shown in green. CMB-based constraints from the Planck 2018 TT,EE,EE+lowl+lowE dataset are shown in yellow (see text for details) and show a clearly complementary degeneracy direction. Constraints from the second and third moments alone (pink) slightly outperform the PDF. We note this is not a contradiction as we use a truncated PDF. Some details on the use of moments for constraints are given in Appendix \ref{app:LDT_moments}.}
  \label{fig:contour_omega_cdm_sigma8_2pcf_pdf}
\end{figure}
\end{centering}

Figure \ref{fig:contour_omega_cdm_sigma8_2pcf_pdf} compares the PDF constraints on $\Omega_{\rm cdm}$ and $\sigma_8$ at two scales to other probes: the two-point correlation function with $\theta_{min}=5^\prime$ and CMB-based constraints from Planck 2018 \citep{Planck18}. The Planck constraints are extracted from the \texttt{TTTEEE\_lowl\_lowE} chains in the Planck legacy archive for the \texttt{base} cosmology. This corresponds to data from the Planck TT, TE and EE spectra at $l>30$ plus low-$l$ TT and EE spectra. In later figures, we use instead the \texttt{base\_w} dataset for the $w$CDM cosmology or \texttt{base\_mnu} for the $\Lambda$CDM+$M_\nu$ cosmology. The covariance matrices are generated specifically using $\Omega_{\rm cdm}/h^2$ as the $\Omega_{\rm cdm}$ value for each run (i.e. not using the values provided for $\Omega_{\rm m}$). It is important to note that all the Planck constraints here come marginalised over the standard $\Lambda$CDM parameters as a minimum (including $h$, $n_s$, $\Omega_{\rm b}$) while our PDF and two-point correlation function Fisher matrices are marginalised only over the parameters shown in each contour plot. 

\begin{centering}
\begin{figure}
\centering
  \includegraphics[width=1\columnwidth]{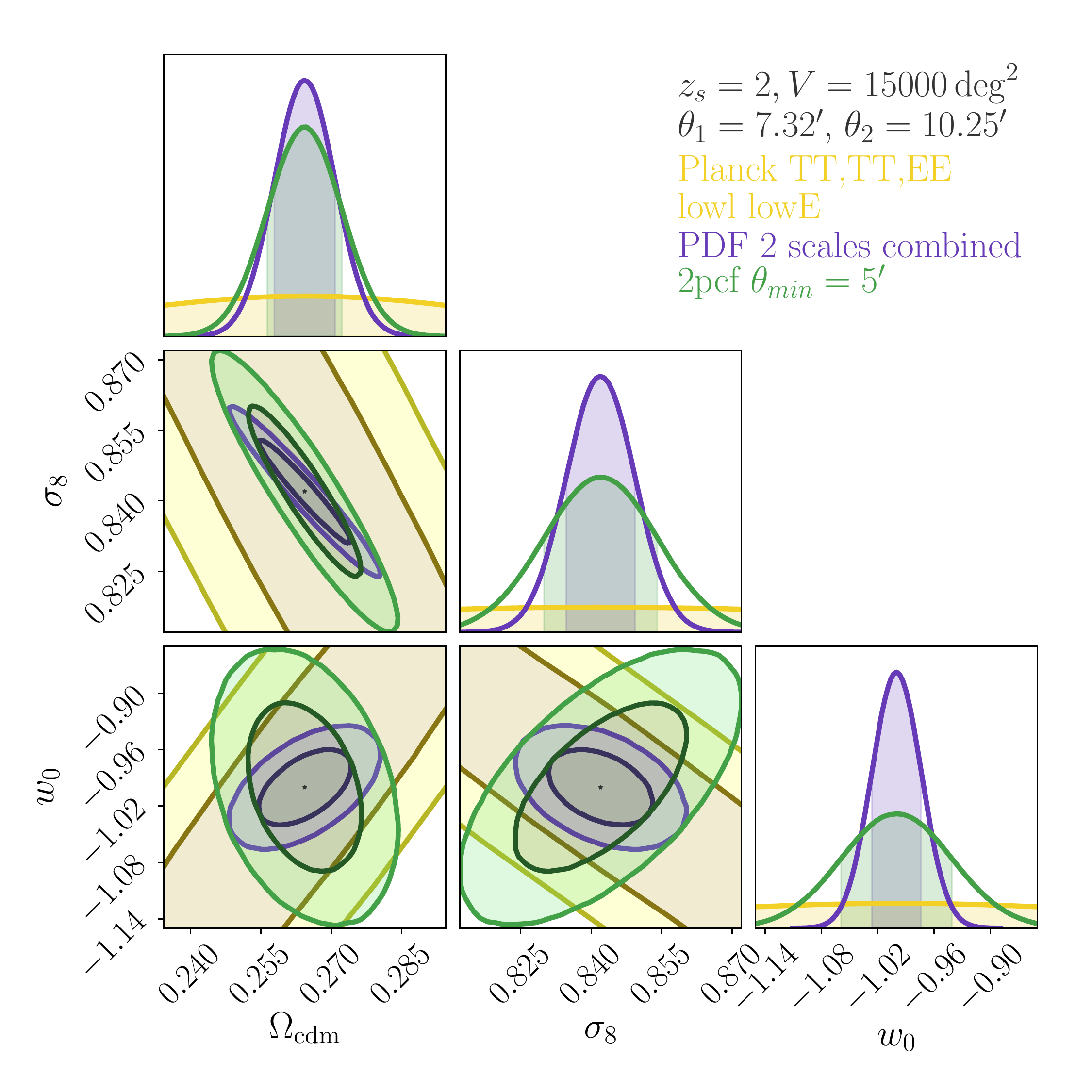}
   \caption{As Figure \ref{fig:contour_omega_cdm_sigma8_2pcf_pdf}, but extended to include the dark energy equation of state $w_0$ as an additional free parameter.}
  \label{fig:contour_w0_2pcf_pdf}
\end{figure}
\end{centering}

\begin{centering}
\begin{figure}
\centering
  \includegraphics[width=1\columnwidth]{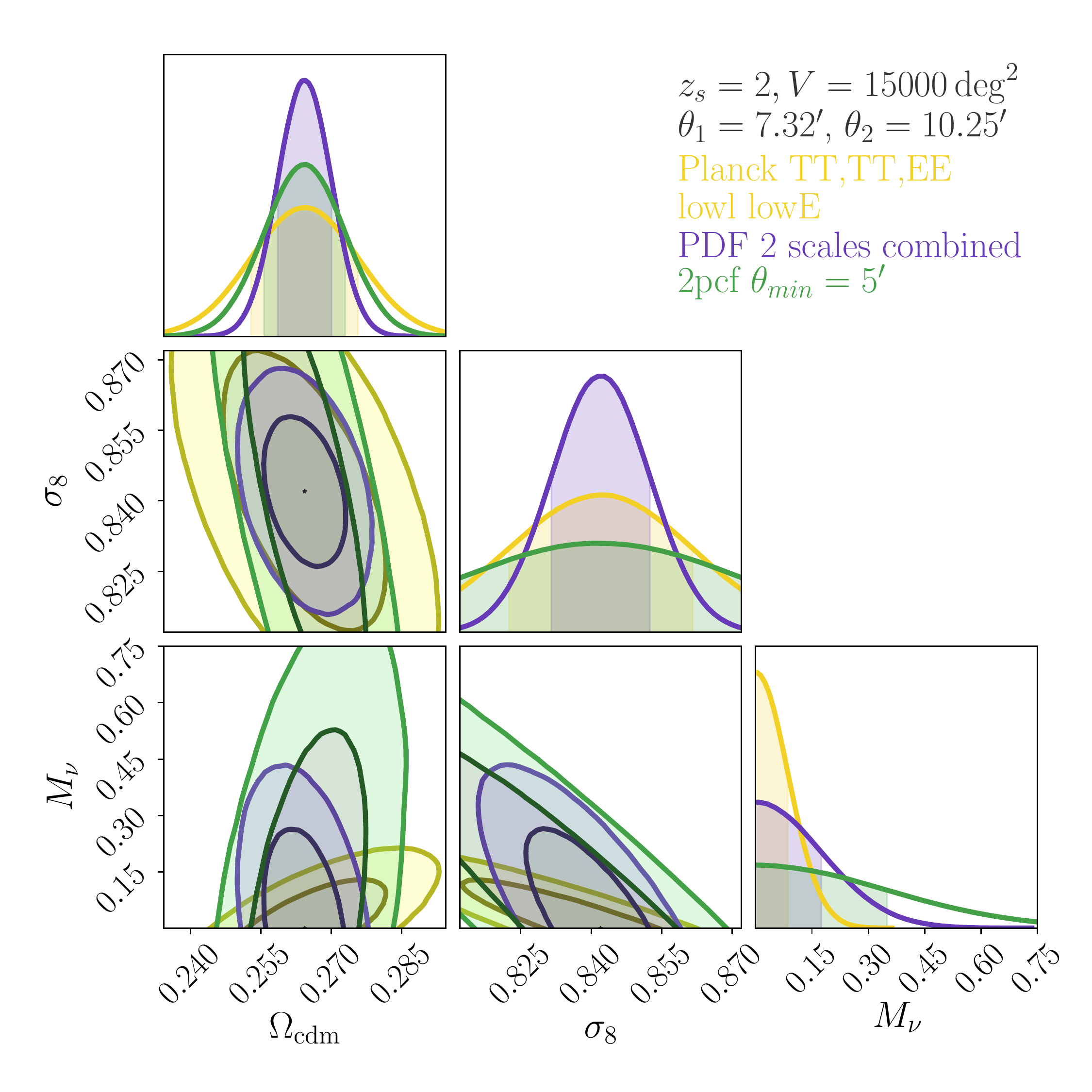}
   \caption{As Figure \ref{fig:contour_omega_cdm_sigma8_2pcf_pdf}, but extended to include the total mass of neutrinos $M_\nu$ as an additional free parameter.}
  \label{fig:contour_mnu_2pcf_pdf}
\end{figure}
\end{centering}

We see that the constraints in Figure \ref{fig:contour_omega_cdm_sigma8_2pcf_pdf} compare favourably with those from the two-point correlation function, although the degeneracy direction in both cases is similar. This is due to the fact that we have fixed all other cosmological parameters here. The degeneracy directions can differ as soon as one varies more parameters. We demonstrate this effect in Figure \ref{fig:contour_w0_2pcf_pdf}, which extends the calculation to include constraints on $w_0$ and looks more promising. Here we show that adding information from the PDF has the potential to break degeneracies between $w_0$ and both $\Omega_{\rm cdm}$ and $\sigma_8$, potentially tightening constraints considerably. We see something similar in Figure \ref{fig:contour_mnu_2pcf_pdf}, which shows simultaneous constraints on $\Omega_{\rm cdm}$, $\sigma_8$ and $M_\nu$. It is not possible to extract meaningful constraints on $M_\nu$ using either CMB or large-scale structure information alone. But the PDF information content compares well with that from the two-point correlation function, and the degeneracy directions are shifted somewhat, allowing for potential degeneracy breaking by combining the two probes. 

We can also use the Fisher matrix forecasting method to analyse the robustness of our results. For instance,we find that implementing the full differential equations for cylindrical collapse instead of the approximation given in Equation \eqref{eq:cylcollapseparam} makes no difference to the results. 

\subsection{Extended Parameter Set Constraints}\label{sec:extended_constraints}

The PDF derivatives for the parameters presented in Figures \ref{fig:contour_omega_cdm_sigma8_scales} to \ref{fig:contour_mnu_2pcf_pdf} have all been validated using simulations (see Section \ref{sec:validation_derivatives}). It is reasonable to assume that it is therefore safe to extend the application of the model to other cosmological parameters to produce more quantitative results relevant for a realistic future surveys. While an in-depth analysis of the relevant degeneracies and a realistic redshift distribution is beyond the scope of this work, we analyse the key effects of varying the full set of cosmological parameters. In Table \ref{tab:constraints}, we present forecasted cosmological constraints from the PDF marginalised over the standard $\Lambda$CDM parameters, and combined with the CMB prior from Planck described in the previous subsection.

As we aim for these results to represent more realistic constraints, we also choose to be more conservative with our shape noise parameters, adopting $n_{\rm gal}$ of 8 arcmin$^{-2}$ instead of 30 arcmin$^{-2}$, which approximately doubles the shape noise standard deviation. We do this because in practice the total number of galaxies would not be located at a single source redshift, but instead distributed across different redshift bins according to the source galaxy distribution $n_s(z_s)$. This can be included in our formalism by replacing the weight function from Equation~\eqref{eq:weight} by
\begin{equation}
\label{eq:weightsource}
    \omega_{n_s}(z) = \frac{3\,\Omega_{\rm m}\,H_0^2}{2\,c^2}\!\! \int\!\! dz_s \frac{d_A(z)\,d_A(z_s,z)}{d_A(z_s)}\, (1+z) n_s(z_s)\,.
\end{equation}

Overall, the constraints in Table \ref{tab:constraints} are very promising. The constraints in the $\Lambda$CDM case are comparable, but the PDF significantly outperforms the two-point correlation function once a free neutrino mass or dark energy equation of state is introduced. We generally find that the resulting contours have quite similar (though not precisely identical) degeneracy directions, as was also seen by \cite{Patton17}, for example. In the $\Lambda$CDM case, strong constraints from Planck mean the advantage of the PDF over the two-point correlation function seen in Figure \ref{fig:contour_omega_cdm_sigma8_2pcf_pdf} becomes less relevant. When additional free parameters that are poorly constrained by the Planck covariance matrix are introduced, this is no longer the case. The PDF can constrain $M_\nu$ over 25\% better and $w_0$ about 40\% better than the two-point correlation function. An examination of the individual Fisher matrices reveals that the advantage of the PDF lies primarily in its ability to better constrain $\sigma_8$ than the two-point correlation function. Of course, both $M_\nu$ and $w_0$ are anti-correlated with $\sigma_8$, resulting in the improved constraints we see in Table \ref{tab:constraints}.

Finally, we note that combining Planck with both the PDF and two-point correlation, we expect a modest improvement of 15\% on $\sigma_8$ constraints for base LCDM. For the extended models we see up to a 50\% improvement for most parameters in the $w_0$CDM and $\nu$LCDM cosmologies, except for $n_s$, which is still strongly constrained by Planck. In particular we noticed a 40-45\% improvement on $\sigma_8$, $M_\nu$, $w_0$ and $h$. Note that the improvement of the PDF over the two-point correlation tends to be reduced by the presence of shape noise. We expect that when repeating a similar analysis at lower source redshifts $z_s\sim 1$ with a larger smoothing scale $\sim 20$ arcmin (thus keeping the degree of non-linearity roughly the same as for our current setup), the impact of shape noise will be smaller and the PDF potentially even more powerful.

\begin{table}
\centering
\begin{tabular}{l|r|r|r}
\hline
\textbf{Planck+PDF vs. Planck} & $\Lambda$CDM & $\Lambda$CDM$+M_\nu$ & $w$CDM\\
\hline
$\Omega_{\rm cdm}$ & +55\% & +43\% & +65\% \\
$\sigma_8$ & +43\% & +40\% & +79\%\\
$\Omega_{\rm b}$ & +52\% & +37\% & +67\%\\
$n_s$ & +25\% & +24\% & +20\%\\
$h$ & +53\% & +39\% & +74\%\\
$M_\nu$ (eV) & - & +31\% & -\\
$w_0$ & - & - & +78\%\\
\hline
\textbf{Planck+PDF vs. Planck+2pcf} & $\Lambda$CDM & $\Lambda$CDM$+M_\nu$ & $w$CDM\\
\hline
$\Omega_{\rm cdm}$ & 0\% & +23\% & +36\% \\
\rowcolor{tablecolour}$\sigma_8$ & +8\% & +32\% & +35\%\\
$\Omega_{\rm b}$ & +0\% & +26\% & +37\%\\
$n_s$ & +0\% & +2\% & +4\%\\
$h$ & +0\% & +25\% & +36\%\\
\rowcolor{tablecolour} $M_\nu$ (eV) & - & +27\% & -\\
\rowcolor{tablecolour} $w_0$ & - & - & +40\%\\
\hline
\end{tabular}
\caption{(Upper panel) Percentage improvements in the 1-$\sigma$ constraints from Planck (\texttt{TTTEEE\_lowl\_lowE}) when adding the convergence PDF at two combined smoothing scales at $z_s=2$ from a Euclid-like survey. (Lower panel) Percentage improvements in the constraints from Planck combined with the PDF over constraints from Planck and the two-point correlation function. Note that we take a more conservative approach to shape noise here compared to in Figures \ref{fig:contour_omega_cdm_sigma8_scales}-\ref{fig:contour_mnu_2pcf_pdf} (see text).}\label{tab:constraints}
\end{table}

\subsection{Comparison with Previous Work}\label{sec:previous_work}

There are several works in the literature that provide us with useful points of comparison for our results. The results of \cite{Patton17} are perhaps the best suited to comparison with the current work. They provide a simulation-based Fisher analysis of the PDF using the fast L-PICOLA perturbative mocks. Although the mocks are based on Lagrangian perturbation theory and not expected to fully capture the non-Gaussian convergence PDF, they confirm that competitive constraints can be achieved from the convergence PDF compared to the cosmic shear power spectrum including an equivalent number of modes in each measurement. They extract their PDFs from a convergence field smoothed on a scale of approximately 13.7 arcmin, which is a little larger than the scales considered for our Fisher forecast. For their Fisher forecast, they vary the parameters \{$\Omega_{\rm m}$, $H_0$, $\sigma_8$\} and include a CMB prior on $\Omega_{\rm m}=\Omega_{\rm m}h^2$. They find that combining the PDF with the power spectrum improves the $\Omega_{\rm m}-\sigma_8$ constraints by a factor of two despite the quite similar degeneracy directions between these two parameters for the two probes. They also highlight that limiting the convergence PDF to values of $\kappa>0$ results in significant degradations in the constraints, and that the convergence PDF provides significant information that is not available in the peaks of the convergence field alone.

\cite{Liu19WLPDF} performed a joint simulation-based analysis of the constraining power of the weak lensing power spectrum and PDF for an Rubin Observatory-like survey, focusing particularly on constraining the neutrino mass. Their simulation-based MCMC forecast (using the same MassiveNuS simulation) also combines information from five source redshift bins. The authors focus on the PDF smoothed in Fourier space, on a scale roughly corresponding to 2 arcmin, which represents a more non-linear regime than that considered in this work, and also give results focused on a single source redshift for $z_s=1$. They also find that the PDF compares well with the power spectrum as a cosmological probe, finding that the PDF provides generally stronger constraints. They find that both combining the PDF and power spectrum and making use of a tomographic analysis can significantly improve constraints. For now we keep these extensions for future work.

A recent paper by \cite{thiele2020accurate} provides a nice complement to the present work. The authors provide an analytical model for the weak lensing convergence PDF based on the halo model, which is suited for smaller smoothing scales but not accurate enough on the mildly non-linear scales considered in this work. The authors provide results from a Fisher forecast for a Rubin Observatory-like survey and show promising results for constraints on the neutrino mass when a CMB prior on $A_s$ is included, and find reasonable agreement with the results of \cite{Liu19WLPDF}.

\citet{Friedrich18} and \citet{Gruen18} indirectly analysed the joint PDF of line-of-sight projected galaxy density and lensing in year-1 data of the Dark Energy Survey, using the technique of density split statistics. In this approach, the PDF of foreground galaxy counts is split into a set of density quantiles, and the strength of tangential gravitational shear around these quantiles is used to probe the underlying projected matter density PDF. Using the modelling of \citet{Friedrich18} (which is based on principles related to the LDT ansatz presented here) \citet{Gruen18} were able to simultaneously measure $\sigma_8$, $\Omega_{\rm m}$, linear galaxy bias and galaxy-matter stochasticity and at the same time test for deviations from the $\Lambda$CDM predictions for the skewness of the matter density field. Their success with observational data is a proof of concept, indicating also that the results presented here carry over to real data analysis.

\section{Conclusion}\label{sec:conclusion}

In this work, we have extended the theoretical model for the weak lensing convergence PDF from \cite{Barthelemy_2020} to include massive neutrinos and the dark energy equation of state. We showed that its percent-level accuracy allows for robust calculations of Fisher derivatives for a range of cosmological parameters through validation with simulated lensing maps, see Figures~\ref{fig:PDFderivatives}-\ref{fig:PDFderivativeMnu}. We found that Halofit \citep{Takahashi_halofit_2012, Bird12} provides sufficiently accurate non-linear variances for the calculation of such derivatives. Furthermore, accurate covariance matrices for Fisher forecasts can be derived from fast simulation codes like FLASK. Therefore, theory-based forecasts as provided in this paper can easily be carried out without the need for expensive N-body/ray-tracing simulations. We have focused on the central region of the PDF, which has a Gaussian likelihood, making the Fisher matrix a natural tool for quantifying the cosmological information contained within it. This is to be contrasted with the first few moments of the convergence field, for which the likelihood of exhibits significant non-Gaussian features (see Figure~\ref{fig:moment_distributions}).

Using our Fisher matrix forecasts, we have demonstrated the value of the weak lensing convergence PDF as a cosmological probe, and its complementarity to the two-point correlation function and CMB data. We have shown that even with a single source redshift, the convergence PDF can provide useful information for constraining cosmology that outperforms that from the convergence two-point correlation function. Our theoretical model has allowed us to perform the first forecast for the weak lensing convergence PDF that varies a complete set of $\Lambda$CDM parameters (see Table \ref{tab:constraints}). Our comparison could be seen as quite conservative, as we focus on a single source redshift, although it is at quite high redshift ($z_s=2$) due to limitations of the smoothing scales accessible in small aperture simulated maps. While lower source redshifts should produce more non-Gaussian information, they also push the model further into the nonlinear regime, which can be compensated by increasing the smoothing scale accordingly. The particular power of the PDF over the two-point correlation comes from its superior capacity to constrain $\sigma_8$ -- this leads directly to better constraints on both $M_\nu$ and $w_0$, both of which are degenerate with $\sigma_8$.

There are many natural extensions to this work. An obvious place to start would be with introducing multiple source redshifts to examine the constraining power of a tomographic analysis, which \cite{Liu_2018} found produced significant improvements in forecasted constraints on $M_\nu$. In \cite{Barthelemy_2020CMB}, the model for the PDF outlined in this work was applied to the CMB lensing PDF, and the cosmological constraints achievable with such a formalism would be another interesting avenue for exploration, although the non-Gaussian information content is smaller so the PDF might be a relatively weaker measurement in this case. 

Finally, we would like to comment on our choice to use an indirect observable, the convergence, rather than shear components for instance, with the risk of introducing 
biases in the conversion. The main advantage of basing a theoretical analysis like this one on weak lensing mass maps is that they allow for easier intuitive understanding of the connection between results for the projected matter density and lensing. Nonetheless, we note that \cite{Patton17} allowed for both multiplicative and additive shear biases in their analysis, and that they found qualitatively similar results with and without marginalisation over these systematics, highlighting that the cosmological information from the PDF seems less sensitive to these biases than the power spectrum. Furthermore, we note that an analysis of one-point statistics for the Dark Energy Survey (DES) has been performed based on moments of the convergence \citep{gatti_dark_2020} and shear profiles around density quantiles \citep{Gruen18, Friedrich18}. We also note that the tangential shear profile around the line of sight can be directly inferred from the convergence profile as in Equation II.6 of \cite{Friedrich18}. \cite{pires_euclid_2020} presented significant improvements on the Kaiser-Squires method with a novel mass-inversion method (KS+) that corrects for systematic effects, and pointed out that convergence maps could play an important role in the analysis for future surveys (highlighting non-Gaussianity studies as a specific example) because the lensing signal is more compressed in convergence maps, leading to less computational expense. We also emphasise once again that the mean (which cannot be reconstructed) has been subtracted in the simulated mass maps we have used. As shown by \cite{Barthelemy_2020} it is possible to extend large deviation theory and the nulling principle to the aperture mass, which is much more tightly linked to measured shear data and we believe that the main conclusion of our investigations will not be changed with such more specific observables.

\section*{Acknowledgements}

This work is partially supported by the SPHERES grant ANR-18-CE31-0009 of the French {\sl Agence Nationale de la Recherche} and by Fondation MERAC.
ABa is supported by a fellowship from CNES. CG acknowledges the grants ASI n.I/023/12/0, ASI-INAF n.  2018-23-HH.0 and PRIN MIUR 2015 Cosmology and Fundamental Physics: illuminating the Dark Universe with Euclid".  CG is also supported by PRIN-MIUR 2017 WSCC32 ``Zooming into dark matter and proto-galaxies with massive lensing clusters''. MB also acknowledges support  by  the  project  ``Combining  Cosmic  Microwave  Back-ground and Large Scale Structure data: an Integrated Approach  for  Addressing  Fundamental  Questions  in  Cosmology",  funded  by  the  PRIN-MIUR 2017  grant 2017YJYZAH.

This work has made use of the Horizon Cluster hosted by Institut d'Astrophysique de Paris. We thank St\'ephane Rouberol for running smoothly this cluster for us. ABa and ABo thank the Institut de Physique Th\'eorique for hosting several visits during the completion of this work. We thank the authors of \cite{Takahashi17} and \cite{Liu_2018} for making their lensing maps publicly available and providing help for analysing their maps. We also thank the Euclid higher-order weak lensing statistics team for useful discussions.

\section*{Data Availability}

There is no new data associated with this article.

\bibliographystyle{mnras}
\bibliography{LSStructure}

\begin{thebibliography}{}
\makeatletter
\relax
\def\mn@urlcharsother{\let\do\@makeother \do\$\do\&\do\#\do\^\do\_\do\%\do\~}
\def\mn@doi{\begingroup\mn@urlcharsother \@ifnextchar [ {\mn@doi@}
  {\mn@doi@[]}}
\def\mn@doi@[#1]#2{\def\@tempa{#1}\ifx\@tempa\@empty \href
  {http://dx.doi.org/#2} {doi:#2}\else \href {http://dx.doi.org/#2} {#1}\fi
  \endgroup}
\def\mn@eprint#1#2{\mn@eprint@#1:#2::\@nil}
\def\mn@eprint@arXiv#1{\href {http://arxiv.org/abs/#1} {{\tt arXiv:#1}}}
\def\mn@eprint@dblp#1{\href {http://dblp.uni-trier.de/rec/bibtex/#1.xml}
  {dblp:#1}}
\def\mn@eprint@#1:#2:#3:#4\@nil{\def\@tempa {#1}\def\@tempb {#2}\def\@tempc
  {#3}\ifx \@tempc \@empty \let \@tempc \@tempb \let \@tempb \@tempa \fi \ifx
  \@tempb \@empty \def\@tempb {arXiv}\fi \@ifundefined
  {mn@eprint@\@tempb}{\@tempb:\@tempc}{\expandafter \expandafter \csname
  mn@eprint@\@tempb\endcsname \expandafter{\@tempc}}}

\bibitem[\protect\citeauthoryear{Amendola et~al.,}{Amendola
  et~al.}{2018}]{Euclid16}
Amendola L.,  et~al., 2018, \mn@doi [Living Reviews in Relativity]
  {10.1007/s41114-017-0010-3}, 21

\bibitem[\protect\citeauthoryear{Barthelemy, Codis  \& Bernardeau}{Barthelemy
  et~al.}{2020a}]{barthelemy2020}
Barthelemy A.,  Codis S.,   Bernardeau F.,  2020a, Probability distribution
  function of the aperture mass field with large deviation theory (\mn@eprint
  {arXiv} {2012.03831})

\bibitem[\protect\citeauthoryear{Barthelemy, Codis, Uhlemann, Bernardeau  \&
  Gavazzi}{Barthelemy et~al.}{2020b}]{Barthelemy_2020}
Barthelemy A.,  Codis S.,  Uhlemann C.,  Bernardeau F.,   Gavazzi R.,  2020b,
  \mn@doi [Monthly Notices of the Royal Astronomical Society]
  {10.1093/mnras/staa053}, 492, 3420–3439

\bibitem[\protect\citeauthoryear{Barthelemy, Codis  \& Bernardeau}{Barthelemy
  et~al.}{2020c}]{Barthelemy_2020CMB}
Barthelemy A.,  Codis S.,   Bernardeau F.,  2020c, \mn@doi [Monthly Notices of
  the Royal Astronomical Society] {10.1093/mnras/staa931}, 494, 3368–3382

\bibitem[\protect\citeauthoryear{{Bernardeau}}{{Bernardeau}}{1992}]{Bernardeau92}
{Bernardeau} F.,  1992, \mn@doi [\apj] {10.1086/171398}, \href
  {http://adsabs.harvard.edu/abs/1992ApJ...392....1B} {392, 1}

\bibitem[\protect\citeauthoryear{{Bernardeau}}{{Bernardeau}}{1994}]{Bernardeau94smoothing}
{Bernardeau} F.,  1994, \aap, \href
  {https://ui.adsabs.harvard.edu/abs/1994A&A...291..697B} {291, 697}

\bibitem[\protect\citeauthoryear{{Bernardeau}}{{Bernardeau}}{1995}]{Bernardeau95ang}
{Bernardeau} F.,  1995, \aap, \href
  {http://adsabs.harvard.edu/cgi-bin/nph-bib_query?bibcode=1995A%26A...301..309B&db_key=AST}
  {301, 309}

\bibitem[\protect\citeauthoryear{Bernardeau \& Reimberg}{Bernardeau \&
  Reimberg}{2016}]{Bernardeau_2016}
Bernardeau F.,  Reimberg P.,  2016, \mn@doi [Physical Review D]
  {10.1103/physrevd.94.063520}, 94

\bibitem[\protect\citeauthoryear{{Bernardeau} \& {Valageas}}{{Bernardeau} \&
  {Valageas}}{2000}]{BernardeauValageas00}
{Bernardeau} F.,  {Valageas} P.,  2000, \aap, \href
  {http://adsabs.harvard.edu/cgi-bin/nph-bib_query?bibcode=2000A%26A...364....1B&db_key=AST}
  {364, 1}

\bibitem[\protect\citeauthoryear{{Bernardeau}, {van Waerbeke}  \&
  {Mellier}}{{Bernardeau} et~al.}{1997}]{1997A&A...322....1B}
{Bernardeau} F.,  {van Waerbeke} L.,   {Mellier} Y.,  1997, \aap, \href
  {https://ui.adsabs.harvard.edu/abs/1997A&A...322....1B} {322, 1}

\bibitem[\protect\citeauthoryear{{Bernardeau}, {Pichon}  \&
  {Codis}}{{Bernardeau} et~al.}{2014a}]{Bernardeau14}
{Bernardeau} F.,  {Pichon} C.,   {Codis} S.,  2014a, \mn@doi [\prd]
  {10.1103/PhysRevD.90.103519}, \href
  {http://adsabs.harvard.edu/abs/2014PhRvD..90j3519B} {90, 103519}

\bibitem[\protect\citeauthoryear{{Bernardeau}, {Nishimichi}  \&
  {Taruya}}{{Bernardeau} et~al.}{2014b}]{BNT}
{Bernardeau} F.,  {Nishimichi} T.,   {Taruya} A.,  2014b, \mn@doi [\mnras]
  {10.1093/mnras/stu1861}, \href
  {https://ui.adsabs.harvard.edu/abs/2014MNRAS.445.1526B} {445, 1526}

\bibitem[\protect\citeauthoryear{{Bird}, {Viel}  \& {Haehnelt}}{{Bird}
  et~al.}{2012}]{Bird12}
{Bird} S.,  {Viel} M.,   {Haehnelt} M.~G.,  2012, \mn@doi [\mnras]
  {10.1111/j.1365-2966.2011.20222.x}, \href
  {https://ui.adsabs.harvard.edu/abs/2012MNRAS.420.2551B} {420, 2551}

\bibitem[\protect\citeauthoryear{Blas, Lesgourgues  \& Tram}{Blas
  et~al.}{2011}]{CLASS}
Blas D.,  Lesgourgues J.,   Tram T.,  2011, \mn@doi [Journal of Cosmology and
  Astroparticle Physics] {10.1088/1475-7516/2011/07/034}, 2011, 034–034

\bibitem[\protect\citeauthoryear{{Brouwer} et~al.,}{{Brouwer}
  et~al.}{2018}]{Brouwer18troughs}
{Brouwer} M.~M.,  et~al., 2018, \mn@doi [\mnras] {10.1093/mnras/sty2589}, \href
  {https://ui.adsabs.harvard.edu/abs/2018MNRAS.481.5189B} {481, 5189}

\bibitem[\protect\citeauthoryear{Clerkin et~al.,}{Clerkin
  et~al.}{2016}]{Clerkin16}
Clerkin L.,  et~al., 2016, \mn@doi [Monthly Notices of the Royal Astronomical
  Society] {10.1093/mnras/stw2106}, 466, 1444–1461

\bibitem[\protect\citeauthoryear{{Codis}, {Pichon}, {Bernardeau}, {Uhlemann}
  \& {Prunet}}{{Codis} et~al.}{2016}]{Codis16b}
{Codis} S.,  {Pichon} C.,  {Bernardeau} F.,  {Uhlemann} C.,   {Prunet} S.,
  2016, \mn@doi [\mnras] {10.1093/mnras/stw1084}, \href
  {http://adsabs.harvard.edu/abs/2016MNRAS.460.1549C} {460, 1549}

\bibitem[\protect\citeauthoryear{{\relax DES Collaboration}}{{\relax DES
  Collaboration}}{2018}]{Abbott_2018}
{\relax DES Collaboration} 2018, \mn@doi [Physical Review D]
  {10.1103/physrevd.98.043526}, 98

\bibitem[\protect\citeauthoryear{Davies, Cautun, Giblin, Li, Harnois-Déraps
  \& Cai}{Davies et~al.}{2020}]{davies2020constraining}
Davies C.~T.,  Cautun M.,  Giblin B.,  Li B.,  Harnois-Déraps J.,   Cai Y.-C.,
   2020, Constraining cosmology with weak lensing voids (\mn@eprint {arXiv}
  {2010.11954})

\bibitem[\protect\citeauthoryear{{Dodelson} \& {Schneider}}{{Dodelson} \&
  {Schneider}}{2013}]{Dodelson_2013}
{Dodelson} S.,  {Schneider} M.~D.,  2013, \mn@doi [\prd]
  {10.1103/PhysRevD.88.063537}, \href
  {https://ui.adsabs.harvard.edu/abs/2013PhRvD..88f3537D} {88, 063537}

\bibitem[\protect\citeauthoryear{{Friedrich} \& {Eifler}}{{Friedrich} \&
  {Eifler}}{2018}]{Friedrich_2018}
{Friedrich} O.,  {Eifler} T.,  2018, \mn@doi [\mnras] {10.1093/mnras/stx2566},
  \href {https://ui.adsabs.harvard.edu/abs/2018MNRAS.473.4150F} {473, 4150}

\bibitem[\protect\citeauthoryear{{Friedrich} et~al.,}{{Friedrich}
  et~al.}{2018}]{Friedrich18}
{Friedrich} O.,  et~al., 2018, \mn@doi [\prd] {10.1103/PhysRevD.98.023508},
  \href {https://ui.adsabs.harvard.edu/abs/2018PhRvD..98b3508F} {98, 023508}

\bibitem[\protect\citeauthoryear{{Friedrich} et~al.,}{{Friedrich}
  et~al.}{2020a}]{Friedrich_in_prep}
{Friedrich} O.,  et~al., 2020a, To be submitted to MNRAS

\bibitem[\protect\citeauthoryear{Friedrich, Uhlemann, Villaescusa-Navarro,
  Baldauf, Manera  \& Nishimichi}{Friedrich et~al.}{2020b}]{Friedrich_2020}
Friedrich O.,  Uhlemann C.,  Villaescusa-Navarro F.,  Baldauf T.,  Manera M.,
  Nishimichi T.,  2020b, \mn@doi [Monthly Notices of the Royal Astronomical
  Society] {10.1093/mnras/staa2160}, 498, 464–483

\bibitem[\protect\citeauthoryear{Gatti et~al.,}{Gatti
  et~al.}{2020}]{gatti_dark_2020}
Gatti M.,  et~al., 2020, \mn@doi [Monthly Notices of the Royal Astronomical
  Society] {10.1093/mnras/staa2680}, 498, 4060

\bibitem[\protect\citeauthoryear{{Giocoli}, {Meneghetti}, {Metcalf}, {Ettori}
  \& {Moscardini}}{{Giocoli} et~al.}{2014}]{Giocoli14}
{Giocoli} C.,  {Meneghetti} M.,  {Metcalf} R.~B.,  {Ettori} S.,   {Moscardini}
  L.,  2014, \mn@doi [\mnras] {10.1093/mnras/stu303}, \href
  {https://ui.adsabs.harvard.edu/abs/2014MNRAS.440.1899G} {440, 1899}

\bibitem[\protect\citeauthoryear{{Giocoli}, {Baldi}  \& {Moscardini}}{{Giocoli}
  et~al.}{2018}]{Giocoli18}
{Giocoli} C.,  {Baldi} M.,   {Moscardini} L.,  2018, \mn@doi [\mnras]
  {10.1093/mnras/sty2465}, \href
  {https://ui.adsabs.harvard.edu/abs/2018MNRAS.481.2813G} {481, 2813}

\bibitem[\protect\citeauthoryear{{Gruen} et~al.,}{{Gruen}
  et~al.}{2018}]{Gruen18}
{Gruen} D.,  et~al., 2018, \mn@doi [\prd] {10.1103/PhysRevD.98.023507}, \href
  {https://ui.adsabs.harvard.edu/abs/2018PhRvD..98b3507G} {98, 023507}

\bibitem[\protect\citeauthoryear{Harnois-Déraps, Martinet, Castro, Dolag,
  Giblin, Heymans, Hildebrandt  \& Xia}{Harnois-Déraps
  et~al.}{2020}]{harnoisderaps2020}
Harnois-Déraps J.,  Martinet N.,  Castro T.,  Dolag K.,  Giblin B.,  Heymans
  C.,  Hildebrandt H.,   Xia Q.,  2020, Cosmic Shear Cosmology Beyond 2-Point
  Statistics: A Combined Peak Count and Correlation Function Analysis of DES-Y1
  (\mn@eprint {arXiv} {2012.02777})

\bibitem[\protect\citeauthoryear{Hartlap, Simon  \& Schneider}{Hartlap
  et~al.}{2006}]{Hartlap06}
Hartlap J.,  Simon P.,   Schneider P.,  2006, \mn@doi [Astronomy \&
  Astrophysics] {10.1051/0004-6361:20066170}, 464, 399–404

\bibitem[\protect\citeauthoryear{Heymans et~al.,}{Heymans
  et~al.}{2020}]{kids2020}
Heymans C.,  et~al., 2020, KiDS-1000 Cosmology: Multi-probe weak gravitational
  lensing and spectroscopic galaxy clustering constraints (\mn@eprint {arXiv}
  {2007.15632})

\bibitem[\protect\citeauthoryear{Hikage et~al.,}{Hikage
  et~al.}{2019}]{HSC_2019}
Hikage C.,  et~al., 2019, \mn@doi [Publications of the Astronomical Society of
  Japan] {10.1093/pasj/psz010}, 71

\bibitem[\protect\citeauthoryear{{Hilbert}, {Hartlap}  \&
  {Schneider}}{{Hilbert} et~al.}{2011}]{Hilbert2011}
{Hilbert} S.,  {Hartlap} J.,   {Schneider} P.,  2011, \mn@doi [Astronomy \&
  Astrophysics] {10.1051/0004-6361/201117294}, \href
  {http://adsabs.harvard.edu/abs/2011A%26A...536A..85H} {536, A85}

\bibitem[\protect\citeauthoryear{Hilbert et~al.,}{Hilbert
  et~al.}{2020}]{Hilbert20}
Hilbert S.,  et~al., 2020, \mn@doi [Monthly Notices of the Royal Astronomical
  Society] {10.1093/mnras/staa281}, 493, 305–319

\bibitem[\protect\citeauthoryear{{Hill} \& {Sherwin}}{{Hill} \&
  {Sherwin}}{2013}]{Hill13}
{Hill} J.~C.,  {Sherwin} B.~D.,  2013, \mn@doi [\prd]
  {10.1103/PhysRevD.87.023527}, \href
  {https://ui.adsabs.harvard.edu/abs/2013PhRvD..87b3527H} {87, 023527}

\bibitem[\protect\citeauthoryear{{Ivanov}, {Kaurov}  \& {Sibiryakov}}{{Ivanov}
  et~al.}{2019}]{Ivanov19}
{Ivanov} M.~M.,  {Kaurov} A.~A.,   {Sibiryakov} S.,  2019, \mn@doi [\jcap]
  {10.1088/1475-7516/2019/03/009}, \href
  {https://ui.adsabs.harvard.edu/abs/2019JCAP...03..009I} {2019, 009}

\bibitem[\protect\citeauthoryear{{Ivezi{\'c}} et~al.,}{{Ivezi{\'c}}
  et~al.}{2019}]{LSST}
{Ivezi{\'c}} {\v{Z}}.,  et~al., 2019, \mn@doi [\apj]
  {10.3847/1538-4357/ab042c}, \href
  {https://ui.adsabs.harvard.edu/abs/2019ApJ...873..111I} {873, 111}

\bibitem[\protect\citeauthoryear{Jarvis, Bernstein  \& Jain}{Jarvis
  et~al.}{2004}]{TreeCorr}
Jarvis M.,  Bernstein G.,   Jain B.,  2004, \mn@doi [Monthly Notices of the
  Royal Astronomical Society] {10.1111/j.1365-2966.2004.07926.x}, 352,
  338–352

\bibitem[\protect\citeauthoryear{{Kaufman}}{{Kaufman}}{1967}]{Kaufman67}
{Kaufman} G.~M.,  1967, Report No. 6710, Center for Operations Research and
  Econometrics, Catholic University of Louvain, Heverlee, Belgium

\bibitem[\protect\citeauthoryear{Kilbinger}{Kilbinger}{2015}]{Kilbinger_2015}
Kilbinger M.,  2015, \mn@doi [Reports on Progress in Physics]
  {10.1088/0034-4885/78/8/086901}, 78, 086901

\bibitem[\protect\citeauthoryear{{Laureijs} et~al.,}{{Laureijs}
  et~al.}{2011}]{Euclid}
{Laureijs} R.,  et~al., 2011, preprint, \href
  {http://adsabs.harvard.edu/abs/2011arXiv1110.3193L} {} (\mn@eprint {arXiv}
  {1110.3193})

\bibitem[\protect\citeauthoryear{Lewis, Challinor  \& Lasenby}{Lewis
  et~al.}{2000}]{CAMB}
Lewis A.,  Challinor A.,   Lasenby A.,  2000, \mn@doi [\apj] {10.1086/309179},
  538, 473

\bibitem[\protect\citeauthoryear{Liu \& Madhavacheril}{Liu \&
  Madhavacheril}{2019}]{Liu19WLPDF}
Liu J.,  Madhavacheril M.~S.,  2019, \mn@doi [Phys. Rev. D]
  {10.1103/PhysRevD.99.083508}, 99, 083508

\bibitem[\protect\citeauthoryear{{Liu}, {Hill}, {Sherwin}, {Petri}, {B{\"o}hm}
  \& {Haiman}}{{Liu} et~al.}{2016}]{Liu16CMBkappaPDF}
{Liu} J.,  {Hill} J.~C.,  {Sherwin} B.~D.,  {Petri} A.,  {B{\"o}hm} V.,
  {Haiman} Z.,  2016, \mn@doi [\prd] {10.1103/PhysRevD.94.103501}, \href
  {https://ui.adsabs.harvard.edu/abs/2016PhRvD..94j3501L} {94, 103501}

\bibitem[\protect\citeauthoryear{Liu, Bird, Matilla, Hill, Haiman,
  Madhavacheril, Petri  \& Spergel}{Liu et~al.}{2018}]{Liu_2018}
Liu J.,  Bird S.,  Matilla J. M.~Z.,  Hill J.~C.,  Haiman Z.,  Madhavacheril
  M.~S.,  Petri A.,   Spergel D.~N.,  2018, \mn@doi [Journal of Cosmology and
  Astroparticle Physics] {10.1088/1475-7516/2018/03/049}, 2018, 049–049

\bibitem[\protect\citeauthoryear{{LoVerde}}{{LoVerde}}{2014}]{LoVerde14}
{LoVerde} M.,  2014, \mn@doi [\prd] {10.1103/PhysRevD.90.083518}, \href
  {http://adsabs.harvard.edu/abs/2014PhRvD..90h3518L} {90, 083518}

\bibitem[\protect\citeauthoryear{{Martinet}, {Harnois-D{\'e}raps}, {Jullo}  \&
  {Schneider}}{{Martinet} et~al.}{2020}]{2020arXiv201007376M}
{Martinet} N.,  {Harnois-D{\'e}raps} J.,  {Jullo} E.,   {Schneider} P.,  2020,
  arXiv e-prints, \href {https://ui.adsabs.harvard.edu/abs/2020arXiv201007376M}
  {p. arXiv:2010.07376}

\bibitem[\protect\citeauthoryear{Mellier}{Mellier}{1999}]{kappadef}
Mellier Y.,  1999, \mn@doi [Annual Review of Astronomy and Astrophysics]
  {10.1146/annurev.astro.37.1.127}, 37, 127

\bibitem[\protect\citeauthoryear{Munshi, Namikawa, McEwen, Kitching  \&
  Bouchet}{Munshi et~al.}{2020a}]{munshi2020morphology}
Munshi D.,  Namikawa T.,  McEwen J.~D.,  Kitching T.~D.,   Bouchet F.~R.,
  2020a, Morphology of Weak Lensing Convergence Maps (\mn@eprint {arXiv}
  {2010.05669})

\bibitem[\protect\citeauthoryear{{Munshi}, {Namikawa}, {Kitching}, {McEwen}  \&
  {Bouchet}}{{Munshi} et~al.}{2020b}]{Munshi_2020}
{Munshi} D.,  {Namikawa} T.,  {Kitching} T.~D.,  {McEwen} J.~D.,   {Bouchet}
  F.~R.,  2020b, \mn@doi [\mnras] {10.1093/mnras/staa2769}, \href
  {https://ui.adsabs.harvard.edu/abs/2020MNRAS.498.6057M} {498, 6057}

\bibitem[\protect\citeauthoryear{{Patton}, {Blazek}, {Honscheid}, {Huff},
  {Melchior}, {Ross}  \& {Suchyta}}{{Patton} et~al.}{2017}]{Patton17}
{Patton} K.,  {Blazek} J.,  {Honscheid} K.,  {Huff} E.,  {Melchior} P.,  {Ross}
  A.~J.,   {Suchyta} E.,  2017, \mn@doi [\mnras] {10.1093/mnras/stx1626}, \href
  {https://ui.adsabs.harvard.edu/abs/2017MNRAS.472..439P} {472, 439}

\bibitem[\protect\citeauthoryear{Peel, Pettorino, Giocoli, Starck  \&
  Baldi}{Peel et~al.}{2018}]{Peel18}
Peel A.,  Pettorino V.,  Giocoli C.,  Starck J.-L.,   Baldi M.,  2018, \mn@doi
  [Astronomy & Astrophysics] {10.1051/0004-6361/201833481}, 619, A38

\bibitem[\protect\citeauthoryear{{Percival} et~al.,}{{Percival}
  et~al.}{2014}]{Percival_2014}
{Percival} W.~J.,  et~al., 2014, \mn@doi [\mnras] {10.1093/mnras/stu112}, \href
  {https://ui.adsabs.harvard.edu/abs/2014MNRAS.439.2531P} {439, 2531}

\bibitem[\protect\citeauthoryear{Petri, Haiman, Hui, May  \& Kratochvil}{Petri
  et~al.}{2013}]{Petri13}
Petri A.,  Haiman Z.,  Hui L.,  May M.,   Kratochvil J.~M.,  2013, \mn@doi
  [Physical Review D] {10.1103/physrevd.88.123002}, 88

\bibitem[\protect\citeauthoryear{Petri, May  \& Haiman}{Petri
  et~al.}{2016}]{Petri_2016}
Petri A.,  May M.,   Haiman Z.,  2016, \mn@doi [Physical Review D]
  {10.1103/physrevd.94.063534}, 94

\bibitem[\protect\citeauthoryear{Pires et~al.,}{Pires
  et~al.}{2020}]{pires_euclid_2020}
Pires S.,  et~al., 2020, \mn@doi [Astronomy \& Astrophysics]
  {10.1051/0004-6361/201936865}, 638, A141

\bibitem[\protect\citeauthoryear{{Puchwein}, {Baldi}  \& {Springel}}{{Puchwein}
  et~al.}{2013}]{Puchwein_Baldi_Springel_2013}
{Puchwein} E.,  {Baldi} M.,   {Springel} V.,  2013, \mn@doi [\mnras]
  {10.1093/mnras/stt1575}, \href
  {https://ui.adsabs.harvard.edu/abs/2013MNRAS.436..348P} {436, 348}

\bibitem[\protect\citeauthoryear{Repp \& Szapudi}{Repp \&
  Szapudi}{2020}]{Repp_2020}
Repp A.,  Szapudi I.,  2020, \mn@doi [Monthly Notices of the Royal Astronomical
  Society: Letters] {10.1093/mnrasl/slaa139}, 498, L125–L129

\bibitem[\protect\citeauthoryear{{Roncarelli}, {Moscardini}, {Borgani}  \&
  {Dolag}}{{Roncarelli} et~al.}{2007}]{Roncarelli07}
{Roncarelli} M.,  {Moscardini} L.,  {Borgani} S.,   {Dolag} K.,  2007, \mn@doi
  [\mnras] {10.1111/j.1365-2966.2007.11914.x}, \href
  {https://ui.adsabs.harvard.edu/abs/2007MNRAS.378.1259R} {378, 1259}

\bibitem[\protect\citeauthoryear{{Sellentin} \& {Heavens}}{{Sellentin} \&
  {Heavens}}{2016}]{Sellentin_2016}
{Sellentin} E.,  {Heavens} A.~F.,  2016, \mn@doi [\mnras]
  {10.1093/mnrasl/slv190}, \href
  {https://ui.adsabs.harvard.edu/abs/2016MNRAS.456L.132S} {456, L132}

\bibitem[\protect\citeauthoryear{Slepian \& Portillo}{Slepian \&
  Portillo}{2018}]{Slepian_2018}
Slepian Z.,  Portillo S. K.~N.,  2018, \mn@doi [Monthly Notices of the Royal
  Astronomical Society] {10.1093/mnras/sty1081}, 478, 516–529

\bibitem[\protect\citeauthoryear{{Szyszkowicz} \&
  {Yanikomeroglu}}{{Szyszkowicz} \& {Yanikomeroglu}}{2009}]{Szyszkowicz2009}
{Szyszkowicz} S.~S.,  {Yanikomeroglu} H.,  2009, \mn@doi [IEEE Transactions on
  Communications] {10.1109/TCOMM.2009.12.070539}, 57, 3538

\bibitem[\protect\citeauthoryear{Takahashi, Sato, Nishimichi, Taruya  \&
  Oguri}{Takahashi et~al.}{2012}]{Takahashi_halofit_2012}
Takahashi R.,  Sato M.,  Nishimichi T.,  Taruya A.,   Oguri M.,  2012, \mn@doi
  [The Astrophysical Journal] {10.1088/0004-637x/761/2/152}, 761, 152

\bibitem[\protect\citeauthoryear{Takahashi, Hamana, Shirasaki, Namikawa,
  Nishimichi, Osato  \& Shiroyama}{Takahashi et~al.}{2017}]{Takahashi17}
Takahashi R.,  Hamana T.,  Shirasaki M.,  Namikawa T.,  Nishimichi T.,  Osato
  K.,   Shiroyama K.,  2017, \mn@doi [The Astrophysical Journal]
  {10.3847/1538-4357/aa943d}, 850, 24

\bibitem[\protect\citeauthoryear{{Taruya}, {Takada}, {Hamana}, {Kayo}  \&
  {Futamase}}{{Taruya} et~al.}{2002}]{Taruya02}
{Taruya} A.,  {Takada} M.,  {Hamana} T.,  {Kayo} I.,   {Futamase} T.,  2002,
  \mn@doi [\apj] {10.1086/340048}, \href
  {http://adsabs.harvard.edu/abs/2002ApJ...571..638T} {571, 638}

\bibitem[\protect\citeauthoryear{Thiele, Hill  \& Smith}{Thiele
  et~al.}{2020}]{thiele2020accurate}
Thiele L.,  Hill J.~C.,   Smith K.~M.,  2020, Accurate Analytic Model for the
  Weak Lensing Convergence One-Point Probability Distribution Function and its
  Auto-Covariance (\mn@eprint {arXiv} {2009.06547})

\bibitem[\protect\citeauthoryear{Touchette}{Touchette}{2011}]{touchette2011}
Touchette H.,  2011, A basic introduction to large deviations: Theory,
  applications, simulations (\mn@eprint {arXiv} {1106.4146})

\bibitem[\protect\citeauthoryear{{Uhlemann}, {Codis}, {Pichon}, {Bernardeau}
  \& {Reimberg}}{{Uhlemann} et~al.}{2016}]{Uhlemann16}
{Uhlemann} C.,  {Codis} S.,  {Pichon} C.,  {Bernardeau} F.,   {Reimberg} P.,
  2016, \mn@doi [\mnras] {10.1093/mnras/stw1074}, \href
  {http://adsabs.harvard.edu/abs/2016MNRAS.460.1529U} {460, 1529}

\bibitem[\protect\citeauthoryear{{Uhlemann}, {Codis}, {Kim}, {Pichon},
  {Bernardeau}, {Pogosyan}, {Park}  \& {L'Huillier}}{{Uhlemann}
  et~al.}{2017}]{Uhlemann17Kaiser}
{Uhlemann} C.,  {Codis} S.,  {Kim} J.,  {Pichon} C.,  {Bernardeau} F.,
  {Pogosyan} D.,  {Park} C.,   {L'Huillier} B.,  2017, \mn@doi [\mnras]
  {10.1093/mnras/stw3221}, \href
  {https://ui.adsabs.harvard.edu/abs/2017MNRAS.466.2067U} {466, 2067}

\bibitem[\protect\citeauthoryear{{Uhlemann}, {Pajer}, {Pichon}, {Nishimichi},
  {Codis}  \& {Bernardeau}}{{Uhlemann} et~al.}{2018a}]{Uhlemann18pNG}
{Uhlemann} C.,  {Pajer} E.,  {Pichon} C.,  {Nishimichi} T.,  {Codis} S.,
  {Bernardeau} F.,  2018a, \mn@doi [\mnras] {10.1093/mnras/stx2623}, \href
  {https://ui.adsabs.harvard.edu/abs/2018MNRAS.474.2853U} {474, 2853}

\bibitem[\protect\citeauthoryear{{Uhlemann}, {Pichon}, {Codis}, {L'Huillier},
  {Kim}, {Bernardeau}, {Park}  \& {Prunet}}{{Uhlemann}
  et~al.}{2018b}]{Uhlemann18cyl}
{Uhlemann} C.,  {Pichon} C.,  {Codis} S.,  {L'Huillier} B.,  {Kim} J.,
  {Bernardeau} F.,  {Park} C.,   {Prunet} S.,  2018b, \mn@doi [\mnras]
  {10.1093/mnras/sty664}, \href
  {https://ui.adsabs.harvard.edu/abs/2018MNRAS.477.2772U} {477, 2772}

\bibitem[\protect\citeauthoryear{Uhlemann, Friedrich, Villaescusa-Navarro,
  Banerjee  \& Codis}{Uhlemann et~al.}{2020}]{Uhlemann_2020}
Uhlemann C.,  Friedrich O.,  Villaescusa-Navarro F.,  Banerjee A.,   Codis S.,
  2020, \mn@doi [Monthly Notices of the Royal Astronomical Society]
  {10.1093/mnras/staa1155}, 495, 4006–4027

\bibitem[\protect\citeauthoryear{{Valageas}}{{Valageas}}{2002a}]{Valageas2002}
{Valageas} P.,  2002a, \mn@doi [\aap] {10.1051/0004-6361:20011663}, \href
  {http://adsabs.harvard.edu/abs/2002A%26A...382..412V} {382, 412}

\bibitem[\protect\citeauthoryear{{Valageas}}{{Valageas}}{2002b}]{Valageas02}
{Valageas} P.,  2002b, \mn@doi [\aap] {10.1051/0004-6361:20011663}, \href
  {http://adsabs.harvard.edu/cgi-bin/nph-bib_query?bibcode=2002A%26A...382..412V&db_key=AST}
  {382, 412}

\bibitem[\protect\citeauthoryear{Vicinanza, Cardone, Maoli, Scaramella  \&
  Er}{Vicinanza et~al.}{2018}]{Vicinanza2018}
Vicinanza M.,  Cardone V.~F.,  Maoli R.,  Scaramella R.,   Er X.,  2018,
  \mn@doi [Physical Review D] {10.1103/physrevd.97.023519}, 97

\bibitem[\protect\citeauthoryear{{Xavier}, {Abdalla}  \& {Joachimi}}{{Xavier}
  et~al.}{2016}]{Xavier2016}
{Xavier} H.~S.,  {Abdalla} F.~B.,   {Joachimi} B.,  2016, \mn@doi [\mnras]
  {10.1093/mnras/stw874}, \href
  {http://adsabs.harvard.edu/abs/2016MNRAS.459.3693X} {459, 3693}

\bibitem[\protect\citeauthoryear{Zürcher, Fluri, Sgier, Kacprzak  \&
  Refregier}{Zürcher et~al.}{2020}]{Zuercher_2020}
Zürcher D.,  Fluri J.,  Sgier R.,  Kacprzak T.,   Refregier A.,  2020,
  Cosmological Forecast for non-Gaussian Statistics in large-scale weak Lensing
  Surveys (\mn@eprint {arXiv} {2006.12506})

\bibitem[\protect\citeauthoryear{\relax Planck~Collaboration}{\relax
  Planck~Collaboration}{2020}]{Planck18}
\relax Planck~Collaboration 2020, \mn@doi [Astronomy & Astrophysics]
  {10.1051/0004-6361/201833910}, 641, A6

\makeatother
\end{thebibliography}

\appendix

\section{Large Deviation Theory and Moments-Based Constraints}\label{app:LDT_moments}

\subsection{Large-deviation theory in a nutshell}
In this appendix, we provide more background equations relevant to the derivation of the theroetical model presented in Section \ref{sec:theoretical_model} and more specific details on the calculation of the matter density CGF in cylinders. For a more elaborate derivation we refer the reader to \cite{Barthelemy_2020}. 

In general terms, the cumulant generating function (CGF) $\phi_X$ of a continuous random variable $X$ (such as the density contrast) is the logarithm of the moment generating function, which is the Laplace transform of the associated PDF

\begin{equation}
    \phi_X(y)=\log(M_X(y))=\sum_{n=1}^{\infty}k_n\frac{y^n}{n!}.
\end{equation}
$k_n$ are the cumulants (connected moments) of the distribution. In our theory, the mean, $k_1$, is assumed to be zero. The scaled cumulant generating function (SCGF), $\varphi_X$,
\begin{equation}
    \varphi_X(y)=\lim_{k_2\rightarrow0}\sum_{n=0}^{\infty}S_n\frac{y^n}{n!}
\end{equation}
is a related quantity that is key to large deviation theory and that defines the reduced cumulants $S_n$
\begin{equation}
\label{eq:redcum}
    S_n=\frac{k_n}{k_2^{n-1}}.
\end{equation}

The CGF and SCGF in the zero variance limit are straightforwardly related as $\varphi_X(y)=\lim_{k_2\rightarrow0}k_2\phi_X(y/k_2)$. For random variables satisfying a large deviation principle, the SCGF is given through Varadhan's theorem as the Legendre-Fenchel transform of $\Psi_X$, which is called the rate function

\begin{equation}
        \varphi_X(y) = \sup_X [yX - \Psi_X(X)].
\end{equation}
When $\Psi$ is convex, this reduces to a simple Legendre transform and

\begin{equation}\label{eq:scgf}
    \varphi_X(y) = yX - \Psi_X(X),
\end{equation}
with the stationary condition defining the relationship between $X$ and $y$

\begin{equation}\label{eq:stationary_cond}
    y=\frac{\partial\Psi_X}{\partial X}.
\end{equation}

The crucial element of LDT is the contraction principle, which gives that the rate function for any variable that is a function of $X$ can be determined assuming only the most likely mapping between the two variables. This is what allows us to define the rate function using only cylindrical collapse in Section \ref{sec:theoretical_model}. The final SCGF of the cylindrically-filtered density is given by

\begin{equation}\label{eq:varphi_cyl}
    \varphi_{\rm cyl}(y) = \sup_\delta\left[y\delta-\frac{\sigma_l^2(D\theta,L,z)}{2\sigma_l^2(D\theta\sqrt{1+\delta},L,z)} \tau^2(1+\delta)\right].
\end{equation}

This can easily be re-written in terms of $\phi_{\rm cyl}(y)$, which is the form that is implemented in our code. This also connects it to the overview given in Section \ref{sec:theoretical_model}, where we deal only in terms of $\phi$ for simplicity.

\subsection{Moments}
In Figure \ref{fig:contour_omega_cdm_sigma8_2pcf_pdf}, we show that a large part of the information contained in the PDF comes from the second and third moments alone. This is to be treated with some caution, however, as we remove information from the PDF by truncating the tails, while the full distribution of values is included in our measurements of the moments. The moments-only constraints also benefit from a smaller Kaufman-Hartlap factor (see Equation \ref{eq:hartlap}). In addition, constraints with a finite number of moments can only perform well when the total number of observables (in the case of Figure \ref{fig:contour_omega_cdm_sigma8_2pcf_pdf}, two scales and two moments, so four) is greater than the number of parameters being constrained. 

\begin{figure}
    \centering
    \includegraphics[width=\columnwidth]{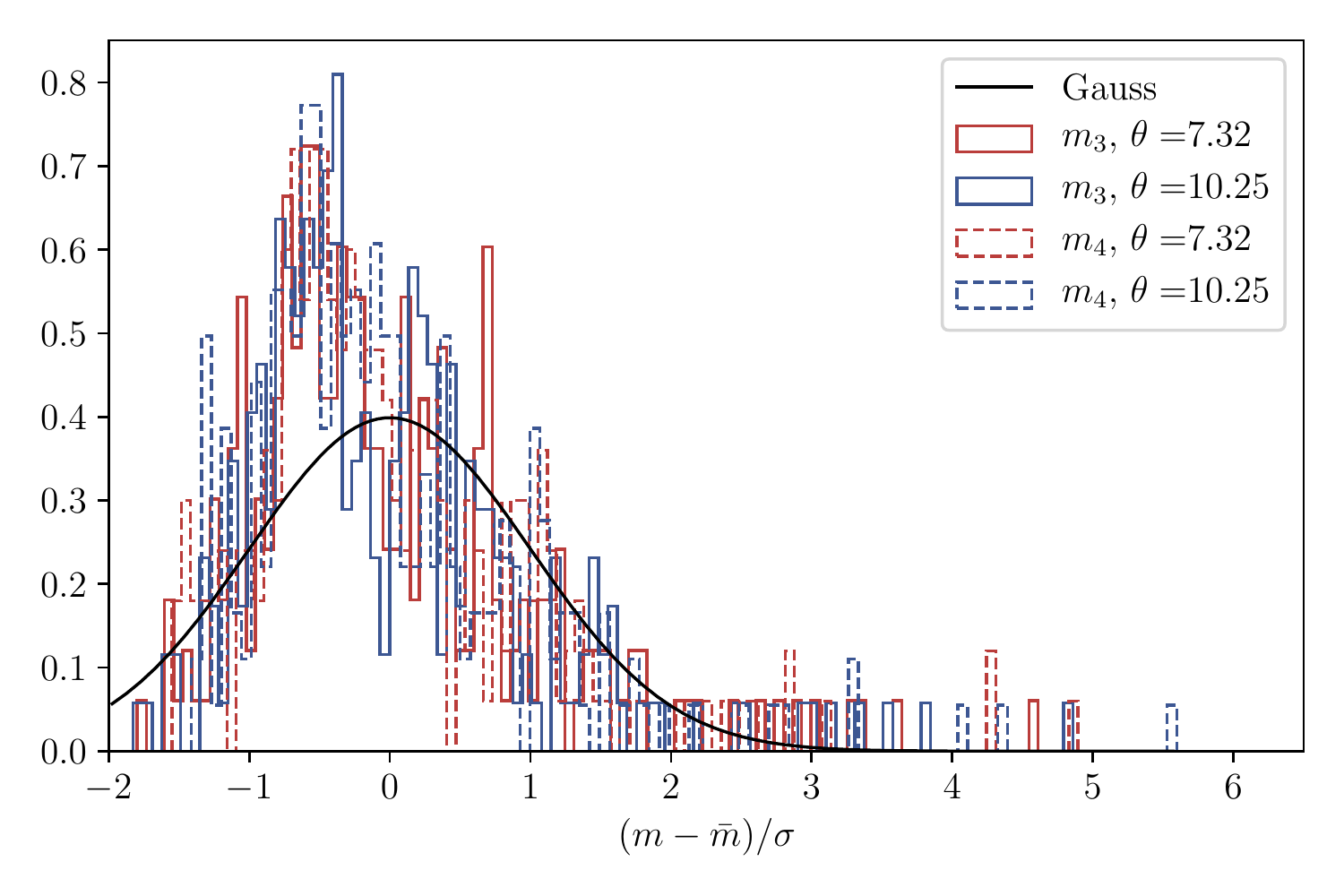}
    \caption{The distributions of the third and fourth central moments taken from 256 25 deg$^2$ patches from FLASK-generated maps.}
    \label{fig:moment_distributions}
\end{figure}

Furthermore, we can highlight a particular advantage of performing Fisher matrix calculations with the bulk of the PDF instead of individual moments: the individual moments show very non-Gaussian distributions when measured from simulations with small patch sizes. Figure \ref{fig:moment_distributions} shows the distributions of the third and fourth moments measured at two smoothing scales measured from patches of FLASK-generated maps with side length 5 deg. This is not the case for the bulk of the PDF as used in this work (see Figure \ref{fig:distributionPDFbins}). 

\section{Simulations and Numerical Tools}\label{sec:sims_and_tools}

\begin{figure*}
    \centering
    \includegraphics[width=2.\columnwidth]{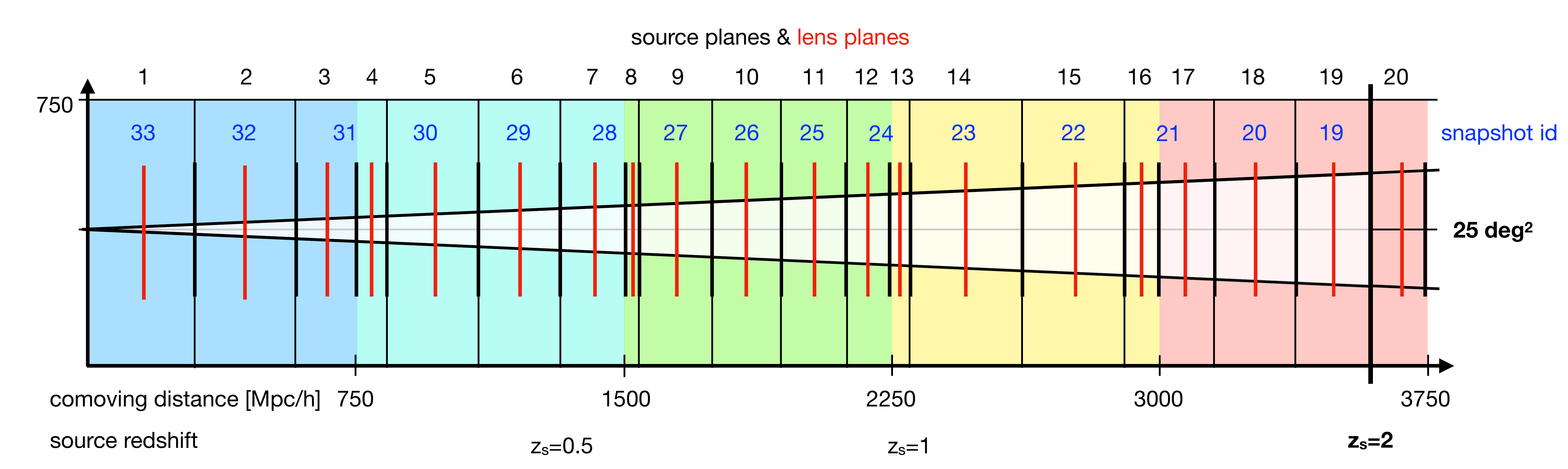}
    \caption{Illustration explaining how the \dustgrainpathfinder~lightcones up to source redshift $z_s=2$ are built from a single realisation N-body simulation of box size 750 Mpc$/h$. To reach the source redshift of $z_s=2$, the box is replicated 5 times along the light-cone up to $z_s=2$. A randomisation procedure is applied to a) avoid replicating the same structure along the line of sight (same colour indicates same randomisation) and b) generate 256 realisations.}
    \label{fig:mapsim}
\end{figure*}

\subsection{\dustgraincosmo}\label{sec:howls}

The \dustgrainpathfinder~simulations \citep[][]{Giocoli18} are a suite of cosmological simulations originally designed to sample several different combinations of modified gravity and massive neutrinos cosmologies -- performed with the {\small MG-GADGET} N-body code \citep[][]{Puchwein_Baldi_Springel_2013} and subsequently post-processed with the weak-lensing map making code {\small MapSim} \citep[][]{Giocoli14}. The original {\small DUSTGRAIN}-{\em pathfinder} suite described in \citet{Giocoli18} has been further extended to include a set of standard $\Lambda $CDM cosmologies with specific deviations of individual cosmological parameters from their fiducial values, as summarised in Table~\ref{tab:cosmology}. We call this new set of simulations the \dustgraincosmo~suite. The simulations follow the evolution of $768^3$ dark matter particles in a periodic cosmological volume of $750$ Mpc/$h$ per side, starting from initial conditions generated at $z=99$ from a random realisation of an initial matter power spectrum computed through the {\small CAMB} Boltzmann solver \citep[][]{CAMB} within the Zel'dovich approximation.

The light-cone is built using the \textsc{Mapsim} routine \citep{Giocoli14} following the geometry illustrated in Figure~\ref{fig:mapsim}. The approach is based on using several snapshots from a single realisation N-body simulation to build a light-cone up to $z=4$. For \dustgraincosmo~there are 21 snapshots available. Given the box length of 750 Mpc$/h$, roughly 7 (5) boxes are needed to cover the comoving distance of about 5 (3.6) Gpc/$h$ to $z_s=4$ ($z_s=2$). To obtain better redshift sampling, each simulation volume necessary for constructing the light-cone is divided along the line-of-sight into multiple redshift slices obtained from the individual snapshots. If the redshift slice reaches beyond the boundary of a single box, two lens planes are constructed from a single snapshot. The total number of lens planes up to $z_s=4$ ($z_s=2$) is 27 (19).
To avoid replicating the same structure along the line of sight, the 7 boxes needed to cover the light-cone are randomised. This randomisation procedure allows to extract multiple realisations from a single simulation.
Randomisation is achieved by using seeds that act on the simulation boxes based on 
\begin{enumerate}
    \item changing the location of the observer, typically placed on the center of one of the faces of the box,
    \item redefining the centre of the box (taking advantage of periodic boundary conditions), and
    \item changing the signs of the box axes.
\end{enumerate}
Note that the same randomisation process is applied to all redshift slices belonging to the same box to avoid spatial discontinuities \citep{Roncarelli07}. 
A comparison of key statistics of the lensing field obtained from different routines based on replicating the original box volume along the line of sight can be found in \cite{Hilbert20}. Note that in the end the mean of the individual weak lensing convergence maps is subtracted.

\subsection{MassiveNuS}\label{sec:MassiveNuS}
The MassiveNuS simulation suite is described in \cite{Liu_2018}. It is based on N-body simulations for dark matter particles with a linear treatment of neutrinos in a box of side length 512 Mpc$/h$.
Similarly to the case for \dustgraincosmo, for a given cosmology, the light-cone is built from a single N-body realisation with ray-tracing as implemented by \texttt{LensTools}. The slices used to construct the light-cone have a constant thickness of 126 Mpc$/h$ and are taken from different randomisations of the single box at the given snapshot. This is different from the {\small MapSim} procedure adopted for \dustgraincosmo, where the same randomisations are kept across different slices fitting in a single box (indicated as a common colour in Figure~\ref{fig:mapsim}). Large sets of maps (10,000 each) usable for validating Fisher derivatives are provided only for two cosmologies, with equal $\Omega_{\rm m}$ and $A_s$ but different values of $M_\nu$ and therefore of $\sigma_8$. Note that the individual $\kappa$ maps in MassiveNuS have non-zero means, which is one difference between these maps and those of \dustgraincosmo.

\subsection{Takahashi full-sky lensing simulations}\label{sec:takahashi}

In Section \ref{sec:validation_fiducial}, we use a PDF measured from the 108 full-sky lensing convergence maps provided by \cite{Takahashi17} to validate our theoretical model. These simulations are ideal for validating our model because they avoid any errors introduced by small patch sizes, but they are only available for a single fiducial cosmology, and so can not be used to validate our cosmological derivatives. The method used to construct these lensing maps differs slightly from that of \dustgraincosmo~and MassiveNuS. The ray-tracing procedure is performed through a set of nested cubic simulation boxes around the observer. Each box has $2048^3$ DM particles, so the resolution is higher in boxes closer to the observer. The final maps are generated with a resolution of 0.86 arcmin. 

\subsection{FLASK}\label{sec:flask}
We use FLASK \citep{Xavier2016} to generate a large number of lensing maps with which to compute the covariance matrix for our PDF Fisher matrix calculations. FLASK can generate lensing convergence maps using two possible approaches. The first is by assuming the convergence field can be well fit by a shifted lognormal distribution. The second approach instead simulates the density field as a function of redshift as a lognormal distribution, and obtains the convergence map by integrating along the line of sight. Here we make use of the first approach which is sufficient to obtain good agreement with the simulation measurements. FLASK is built to generate lensing convergence (or other) maps whose one-point distribution is close to a shifted lognormal \citep{Clerkin16,Hilbert2011}
\begin{equation}
    \label{eq:lognormalPDF}
\mathcal P_{\rm LN}(\kappa) = \frac{\Theta(\kappa-\kappa_0)}{(\kappa-\kappa_0)\sqrt{2\pi}\sigma}\exp\left[-\frac{\left(\ln(\kappa-\kappa_0)-\mu\right)^2}{2\sigma^2}\right]\,,
\end{equation}
where the mean is $\mu=\ln(-\kappa_0+\langle\kappa\rangle)-\sigma^2/2$. The variance of $\kappa$ is related to the parameters as $\sigma_\kappa^2=\exp(2\mu+\sigma^2)(\exp(\sigma^2)-1)$. 

The primary input required by FLASK is a set of lensing $C_l$ generated using a Boltzmann solver like CLASS or CAMB, which we generate up to $l=10,000$. The shift parameter, $\lambda$, is then set to replicate the desired skewness \citep{Xavier2016}

\begin{align}\label{eq:xavier_lambda}
    \lambda &= \frac{\sigma}{\tilde{\mu}_3}\left(1+y(\tilde{\mu}_3)^{-1}+y(\tilde{\mu}_3)\right)-\langle \kappa\rangle, \\
    y(\tilde{\mu}_3) &= \sqrt[3]{\frac{2+\tilde{\mu}_3^2+\tilde{\mu}_3\sqrt{4+\tilde{\mu}_3^2}}{2}}\,, \notag
\end{align}
where $\langle \kappa\rangle$ is the desired mean, $\sigma$ the target variance and $\tilde \mu_3$ the skewness. We use FLASK to generate maps of \textsc{healpix} resolution $N_{\rm side}=4096$, which corresponds to a pixel size of approximately 0.86 arcmin, and then smooth to our required scales. In the end, we find that the mean FLASK PDF agrees with the theoretical one to within 1.25\% for all bins used in our Fisher analysis for both smoothing scales. 
 Note that our procedure relies on a theorem by \citet{Szyszkowicz2009}. They show that the average of correlated lognormal random variables yields a random variable well described by a lognormal PDF with the same value of $\tilde{\mu}_3$. This allows us to impose our desired value of the skewness on the grid scale and obtain the same value also on larger smoothing scales.

\subsection{Limitations of measurements in small simulated maps}
\label{app:patches}

\begin{centering}
\begin{figure}
    \centering
    \includegraphics[width=\columnwidth]{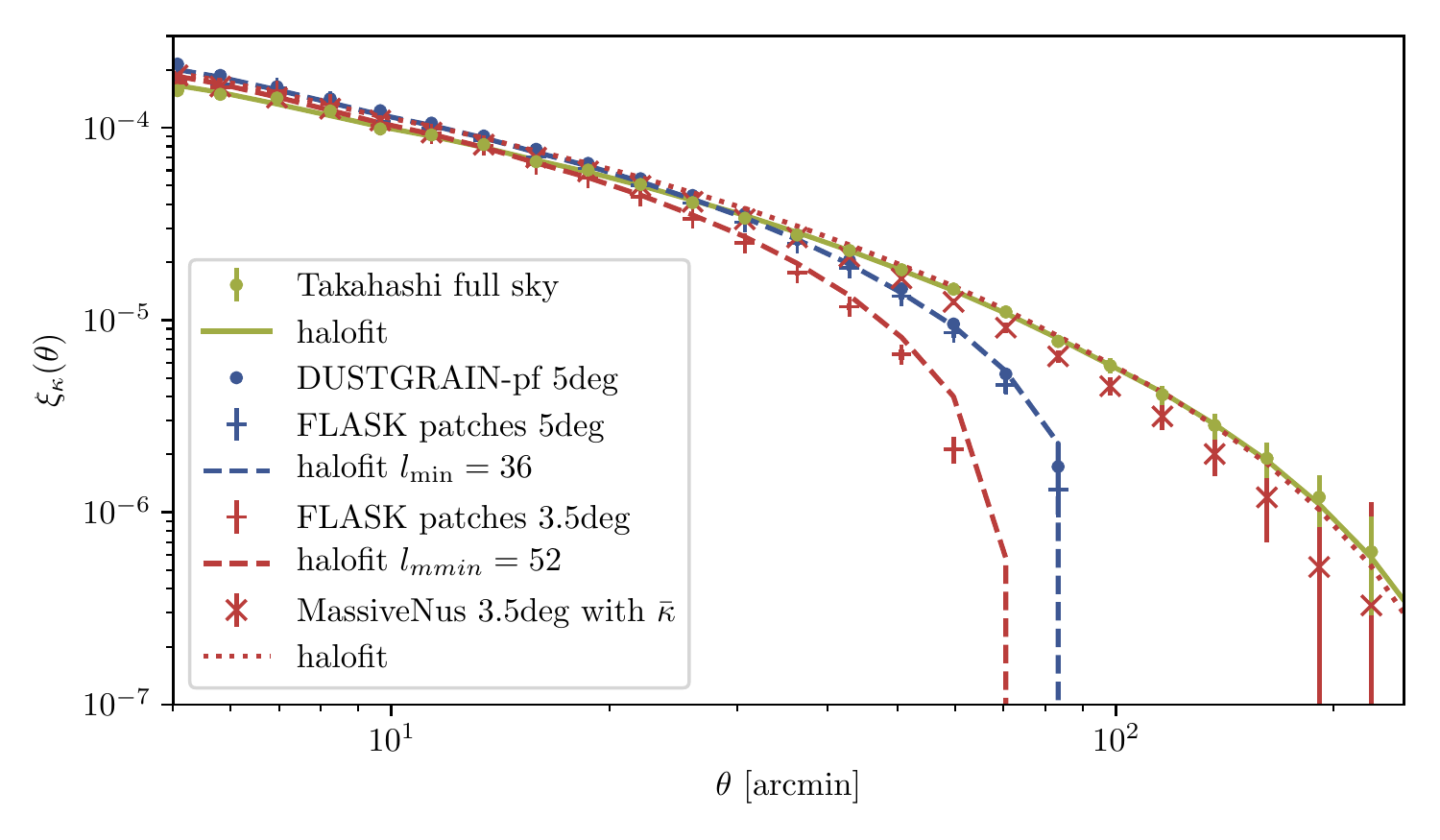}
    \includegraphics[width=\columnwidth]{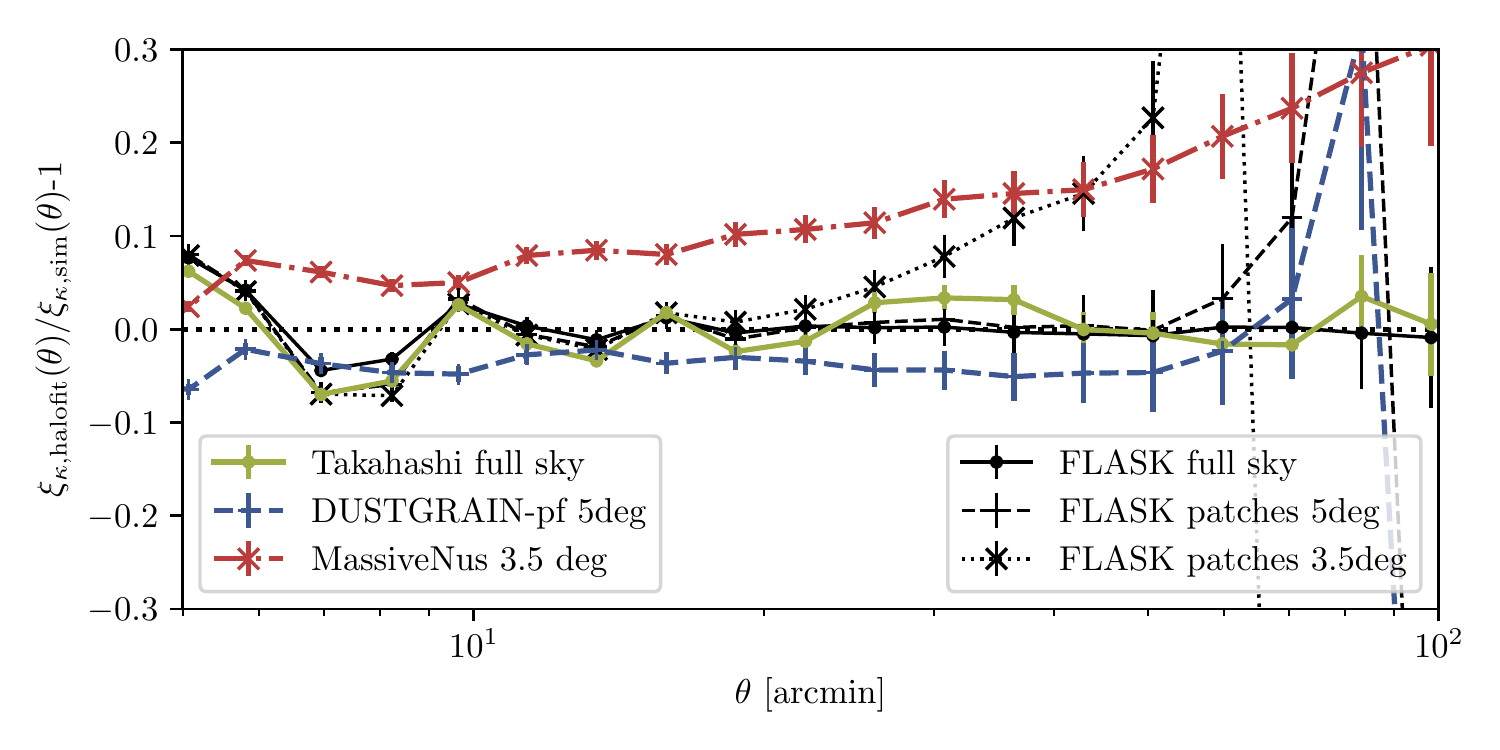}
    \caption{(Upper panel) Logarithmic view of the $\kappa$ correlation functions measured from different simulations (data points). The lines show the halofit predictions for full sky (solid and dotted line) and small patches of 5 and 3.5 deg with subtracted mean (dashed lines). To show the cutoff that is imposed by using small patches with subtracted mean $\bar\kappa$, we plot measurements from small patches cut from the FLASK full sky maps which have been generated with the halofit power spectrum. MassiveNuS maps have been obtained from maps including the mean $\bar\kappa$ instead which restores large-scale power. (Lower panel) Residuals between the curves shown above.}
    \label{fig:2pcf_sims_halofit}
\end{figure}
\end{centering}

In Figure~\ref{fig:2pcf_sims_halofit} we show a comparison of the measured two-point correlation function in the Takahashi full-sky maps \citep{Takahashi17}, as well as the small patches from the \dustgrainpathfinder~simulations \citep{Giocoli18} and MassiveNuS \citep{Liu19WLPDF} to the Halofit prediction \citep{Takahashi_halofit_2012, Bird12} as implemented in CLASS. Note that the convergence maps from the \dustgrainpathfinder~ simulations have been constructed to remove the mean $\bar \kappa$, while this mean is included in the MassiveNuS maps. The upper panel shows an overall good agreement between the measurements and the predictions when taking into account that a small map of side length $L$ with mean subtraction requires to perform a cut in the l-range with roughly $l_{\rm min}\simeq 180 \text{deg}/L$. 
The lower panel focuses on residuals and shows an exquisite agreement between Halofit and the full-sky maps from \citep{Takahashi17}, but a few-percent discrepancies between Halofit and the two-point correlation functions measured from small maps of \dustgrainpathfinder~and MassiveNuS. To establish a baseline and isolate potential errors coming from the finite range of $l\in [l_{\rm min},l_{\rm max}=20000]$, we also show residuals for $\kappa$ maps from FLASK (generated from the Halofit $C_l$s) when considering a full sky or smaller patches (black lines). Note that the residuals seen in Figure \ref{fig:2pcf_sims_halofit} are very consistent across all \dustgraincosmo
~cosmologies, which ultimately means that they cancel out in the derivatives and we find very good agreement between simulation and theory.

\begin{centering}
\begin{figure}
    \centering
    \includegraphics[width=\columnwidth]{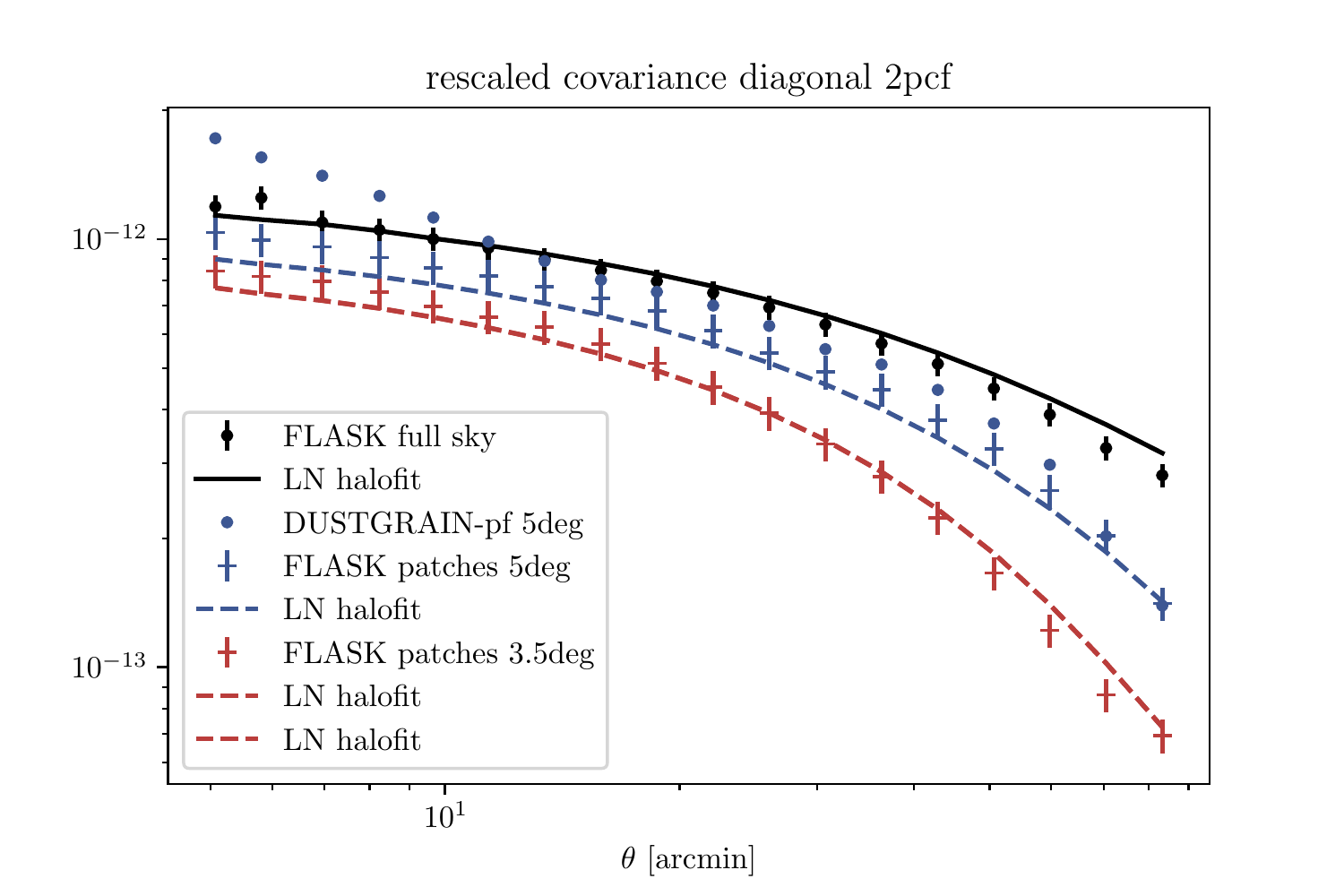}
    \caption{Variance of the $\kappa$ two-point correlation function bins rescaled to a common sky area of 15000 deg$^2$. 
    We validate the theoretical prediction from the lognormal model (black line) against the measurements from full sky FLASK maps (black data points). We also show that considering small patches with mean subtraction (blue and red crosses) underestimate the covariance and are equivalent to imposing an $l_{\rm min}$ cut appropriate for the different patch sizes in the theory (dashed lines). Those predictions are in broad agreement with the measurements from small patch simulations of \dustgrainpathfinder~ (blue points), with  deviations driven by residuals of the two-point correlation with respect to Halofit (see Figure~\ref{fig:2pcf_sims_halofit}).}
    \label{fig:2pcf_sims_halofit_covariance}
\end{figure}
\end{centering}

In Figure~\ref{fig:2pcf_sims_halofit_covariance} we demonstrate that the covariance of the two-point correlation measured from small patches significantly underestimates the covariance from full-sky maps. This can artificially improve the constraining power of the two-point correlation function (or equivalently the power spectrum $C_l$) compared to higher-order statistics, which can be less affected from small patch sizes. As we demonstrated in the main text, the covariance of the $\kappa$ PDF bins are largely independent of the patch size. Hence we caution that analyses based on covariances from small patches can  underestimate the constraining power of the convergence PDF compared to the two-point correlation.

\section{Performing the Laplace Transform using an Effective Mapping}\label{app:effective_mapping}

To perform the inverse Laplace transform in Equation \ref{eq:laplace_trans}, one can apply a method first described in \cite{BernardeauValageas00}. We assume the difference between the unsmoothed Gaussian initial field and the evolved, smoothed field can be combined into a single effective mapping, $\zeta(\tau_{\rm eff})$. In this case, the SCGF (see Equation \ref{eq:scgf}) can be written as

\begin{equation}\label{eq:effective_mapping}
    \varphi(y) = y\zeta(\tau_{\rm eff})-\frac{1}{2}\tau_{\rm eff}^2.
\end{equation}
The stationary condition (Equation \ref{eq:stationary_cond}) then becomes

\begin{equation}
    y=\frac{\textrm{d}}{\textrm{d}\zeta}\frac{\tau_{\rm eff}^2}{2} = \tau_{\rm eff}\left(\frac{\textrm{d}\zeta(\tau_{\rm eff})}{\textrm{d}\tau_{\rm eff}}\right)^{-1}.
\end{equation}
The effective mapping $\zeta(\tau_{\rm eff})$ can now be written in the form of a vertex generating function

\begin{equation}
    \zeta(\tau_{\rm eff})=\sum_{k=0}^n \frac{\mu_k}{k!}\tau_{\rm eff}^k,
\end{equation}
where $\mu_0=0$, $\mu_1=1$. The remaining coefficients can be fitted to the derived SCGF we want to transform, and this fit can then easily be continued into the complex plane. To fit the values of $\mu$, we must express both $\zeta(\tau_{\rm eff})$ and $\tau_{\rm eff}$ in terms of the variable $y$. This can be achieved using Equation \ref{eq:scgf}
\begin{equation}
    \frac{\textrm{d}\varphi(y)}{\textrm{d}y} = \zeta(\tau_{\rm eff}),
\end{equation}
and from Equation \ref{eq:effective_mapping} therefore

\begin{equation}
    \frac{1}{2}\tau_{\rm eff}^2= y\frac{\textrm{d}\varphi(y)}{\textrm{d}y} - \varphi(y).
\end{equation}

We find that the fitting of $\mu$ using a table of $\tau_{\rm eff}$ and $\zeta(\tau_{\rm eff})$ values performs well with a polynomial of order 9. In practice one can also perform the fit for the CGF $\phi(y)$ instead, with the $\mu$ values being linked as $\mu_k^{\rm CGF}=\sigma^k\mu^k$. Ultimately, this allows for straightforward numerical computation of the inverse Laplace transform.

Technically the convergence itself does not satisfy a large deviation principle, unlike the projected density in slices used to build up the model used here. However, this effective mapping approach still works well for the convergence because the form of the bulk of the PDF is dominated by the first few cumulants and its unimodality, while the assumption of LDT only affects the PDF around the critical point quite far into the tails. \cite{barthelemy2020} tried applying the effective mapping approach to invididual redshift slices, but found the result was not significantly different to that obtained by applying the effective mapping approach directly to the convergence CGF.

\end{document}